\documentclass[journal=jacsat,manuscript=article]{achemso}

\usepackage[version=3]{mhchem} 
\usepackage{amsmath,amssymb}
\usepackage[utf8]{inputenc}
\usepackage[T1]{fontenc} 
\usepackage{mathptmx}
\usepackage{bm}
\usepackage{etoolbox}
\usepackage[labelsep=period]{caption}
\usepackage{dcolumn}
\usepackage{multirow}
\usepackage{threeparttable}
\usepackage{algorithm,algorithmic}
\usepackage{mdframed}
\usepackage{tikz}
\usetikzlibrary{shapes.geometric, arrows}
\usepackage{subfig}
\usepackage{graphicx}
\usepackage{tabularx}
\usepackage{array}
\usepackage{booktabs}
\usepackage[section]{placeins}
\usepackage{siunitx}

\newcommand{\pint}[2]{\left(#1 \vert #2\right)}

\newcolumntype{Y}{>{\centering\arraybackslash}X}
\newcolumntype{Z}{>{\raggedleft\arraybackslash}X}
\newcolumntype{W}{>{\raggedright\arraybackslash}X}

\DeclareSIUnit\angstrom{\text {Å}}
\DeclareSIUnit\wn{\cm\tothe{-1}}

\definecolor{revisions}{rgb}{0.0,0.0,0.0}

\SectionNumbersOn

\author{Christopher L. Malbon}
\affiliation[Princeton University]{Department of Chemistry, Princeton Univeristy, Princeton, NJ 08544}
\author{Sharon Hammes-Schiffer}
\affiliation[Princeton University]{Department of Chemistry, Princeton Univeristy, Princeton, NJ 08544}
\email{shs566@princeton.edu}

\title[Multiref NEO 1]
  {Nuclear-Electronic Orbital Multireference Configuration Interaction for Ground and Excited Vibronic States \color{revisions} and Fundamental Insights into Multicomponent Basis Sets}

\abbreviations{}

\begin{document}
%
%

\begin{abstract}
The nuclear-electronic orbital (NEO) approach incorporates nuclear quantum effects into quantum chemistry calculations by treating specified nuclei quantum mechanically, equivalently to the electrons.
Within the NEO framework, excited states are vibronic states representing electronic and nuclear vibrational excitations.
The NEO multireference configuration interaction (MRCI) method presented herein provides accurate ground and excited vibronic states.
The electronic and nuclear orbitals are optimized with a NEO multiconfigurational self-consistent field (NEO-MCSCF) procedure, thereby including both static and dynamic correlation and allowing the description of double and higher excitations.
The accuracy of the NEO-MRCI method is illustrated by computing the ground state protonic densities and excitation energies of the vibronic states for four molecular systems with the hydrogen nucleus treated quantum mechanically.
In addition, revised conventional electronic basis sets adapted for quantized nuclei are developed and shown to be essential for achieving this level of accuracy.
The NEO-MRCI approach, as well as the strategy for revising electronic basis sets, will play a critical role in multicomponent quantum chemistry.
\end{abstract}

\section{Introduction} \label{sec:intro}
The Born-Oppenheimer (BO) approximation entails the separation of the light and fast electrons from the heavy and slow nuclei to solve the time-independent Schr\"odinger equation as a tractable, two step problem. 
The first step of the approximation solves the electronic structure of a static nuclear configuration and introduces the concept of a potential energy surface (PES), where nuclear motion is guided by the potential created by the electrons.
The second step is solving for the nuclear motion on the electronic PES, locating transition states and equilibrium structures that are central to our understanding of chemical reactions or performing molecular dynamics simulations.\cite{Tully2012}
The utility of the BO approximation is widespread. 
However, for many systems and chemical phenomena, invoking this approximation requires additional steps to correct for errors arising from the classical treatment of the nuclei.
For example, zero-point energy and hydrogen tunneling corrections can be applied to calculations of relative energies and rate constants.

Multicomponent methods treat specified nuclei as quantum mechanical wavefunctions by solving a mixed nuclear-electronic Schr\"odinger equation.\cite{PavosevicCulpittSHS2020}
In these methods, the quantum system includes the electrons and the selected nuclei, typically protons, and the electronic structure calculation now incorporates the quantum effects of the nuclei, such as zero point energy\cite{TuckermanMarxKleinParrinello1997,RaugeiKlein2003} and tunneling.\cite{ChaMurrayKlinman1989} 
The remaining classical nuclei parameterize the multicomponent PES, which is now a nuclear-electronic vibronic PES. 
For a predominantly electronically adiabatic system, the lower excited vibronic surfaces correspond to vibrational excitations of the quantum nuclei, whereas the higher excited vibronic surfaces represent electronic excitations or mixed electronic-vibrational excitations. 

Nuclear-electronic orbital (NEO) theory is a well-established framework for multicomponent quantum chemistry.\cite{WebbIordanovSHS2002,SHS2021}
To take advantage of existing, well-tested integral and basis set codes, quantum nuclei are represented by nuclear orbitals expanded in a Gaussian basis set. \cite{YuPavosevicSHS2020}
The NEO Hartree-Fock (NEO-HF) wavefunction is the product of an electronic and a nuclear Slater determinant composed of electronic and nuclear orbitals, respectively.
A NEO multiconfigurational wavefunction is defined as a linear combination  of such products, and a NEO active space is defined by the number of electrons and electronic orbitals and the number of quantum nuclei and nuclear orbitals.
The NEO approach has been successfully implemented for many existing electronic structure methods, including but not limited to density functional theory (DFT),\cite{PakChakrabortySHS2007,ChakrabortyPakSHS2008} time-dependent density functional theory (TDDFT),\cite{YangCulpittSHS2018, CulpittYangPavosevicTaoSHS2019} time-dependent Hartree-Fock (TDHF)\cite{YangCulpittSHS2018}, many-body perturbation theory,\cite{SwalinaPakSHS2005,FajenBrorsen2021mp4,HaseckeMata2024} multistate density functional theory,\cite{YuSHS2020,DickinsonYuSHS2023} coupled cluster theory, \cite{PavosevicCulpittSHS2019,FowlerBrorsen2022,PavosevicSHS2022,PavosevicTaoCulpittZhaoLiSHS2020} complete active space self-consistent field (CASSCF),\cite{WebbIordanovSHS2002,Brorsen2020} nonorthogonal configuration interaction,\cite{SkonePakSHS2005} density matrix renormalization group (DMRG),\cite{MuoloBaiardiFeldmannReiher2020} and time-dependent configuration interaction (TDCI).\cite{GarnerUpadhyayLiSHS2024} 

Computing quantitatively accurate excited vibronic states is a formidable challenge for multicomponent methods.
Multicomponent linear response methods such as NEO-TDHF and NEO-TDDFT have reported accurate proton vibrational excitation energies.\cite{CulpittYangPavosevicTaoSHS2019} 
In such calculations, however, the ground state protonic density is described by NEO-HF, where it is much too localized, or NEO-DFT, where it is improved with the epc17 electron-proton correlation functionals\cite{BrorsenYangSHS2017,YangBrorsenCulpittPakSHS2017} but is still too localized.\cite{PavosevicCulpittSHS2020}
Moreover, linear response methods based on the underlying adiabatic approximation cannot accurately describe multiple excitations, such as double excitations (i.e., a simultaneous electronic and protonic excitation or a double electronic excitation).
In addition, the accurate description of hydrogen tunneling systems within the multicomponent framework requires multireference methods.\cite{PakSHS2004,YuSHS2020,DickinsonYuSHS2023}

These challenges can be overcome by multicomponent multiconfigurational approaches. 
All multicomponent multiconfigurational methods recover some of the electron-proton dynamic correlation shown to be critical in describing nuclear quantum effects.\cite{PakSHS2004,PavosevicCulpittSHS2020,FajenBrorsen2021,PavosevicSHS2022}
However, when static correlation is important, as for hydrogen tunneling systems, multiconfigurational methods based on a single reference fail to provide even qualitatively accurate results.\cite{PakSHS2004}
Multicomponent multireference methods, such as multiconfigurational self-consistent field (MCSCF), describe such systems well, \cite{WebbIordanovSHS2002,PakSHS2004} but the exponential growth of the MCSCF wavefunction limits the size of the active space, thereby omitting sufficient dynamic correlation required for quantitative accuracy.\cite{Lischka2018}
Larger active spaces can be used in MCSCF calculations with heat-bath CI\cite{SmithMussardHolmesSharma2017} or restricted active space methods,\cite{MaLiManniGagliardi2011} as well as in DMRG calculations.\cite{WhiteMartin1999}
Although promising, multicomponent applications of these methods have been limited.\cite{YangWhite2019,Brorsen2020,MuoloBaiardiFeldmannReiher2020,FajenBrorsen2021,FeldmannMuoloBaiardiReiher2022}
Alternatively, electron-proton dynamic correlation can be included post-MCSCF via perturbation theory, as in conventional electronic structure theory,\cite{AnderssonMalmqvistRoosSadlejWolinski1990,AnderssonMalmqvistRoos1992,RoosAnderssonFulscherSerranoPierlootMerchanMolina1996,SharmaBaoTruhlarGagliardi2021} or CI expansions based on an active space reference.

This work presents a multicomponent multireference configuration interaction (NEO-MRCI) approach, which uses  electronic and nuclear orbitals optimized with NEO-MCSCF to produce an MRCI wavefunction.
In conventional quantum chemistry, MRCI represents the highest-level treatment of excited electronic states, particularly for problems in photochemistry and nonadiabatic dynamics.\cite{Yarkony2012,Matsika2021}
As part of our implementation of the NEO-MRCI method, we have created a suite of multicomponent multireference methods.
This suite includes NEO-MCSCF with either a complete active space (NEO-CASSCF) or a restricted active space (NEO-RASSCF), NEO-CI with single excitations or single and double excitations (NEO-SCI and NEO-SDCI), and full CI (NEO-FCI). Our applications of the NEO-MRCI method to several small molecular systems illustrate that the ground state protonic densities and low-lying excited vibronic state excitation energies are significantly more accurate than any previous multicomponent quantum chemistry method to date.

An important conceptual advance of this work is the revision of electronic basis sets for multicomponent calculations.
Conventional electronic basis sets have been developed within the BO approximation, where the nuclei are represented as stationary point charges. Such basis sets are not expected to be suitable for multicomponent calculations, where the nucleus is described as a delocalized nuclear density, for two reasons.
First, the 1s orbital of a conventional electronic basis set is formed by several Gaussian primitives tightly contracted to model the cusp of a Slater-type orbital,\cite{Kato1957} {\color{revisions}but there should be no single electronic-nuclear cusp if the nucleus is not treated as a stationary point charge.
For a delocalized nuclear density, there will be many points at which the electronic and nuclear densities coincide, and treating one point differently from all other points is unbalanced. 
Second, if} the nucleus is treated quantum mechanically, the electronic orbitals will be more diffuse because the electrons are coupled to the delocalized nuclear density.
Revised basis sets, such as those developed and used in this work, are expected to improve the accuracy of all multicomponent methods. 

The paper is organized as follows. 
Section \ref{sec:theory_and_comp_meth} presents the NEO-MRCI formalism, providing an overview of the NEO-MCSCF implementation and reviewing the NEO-CI method.
The revised electronic basis set used in this work is presented in Section \ref{subsec:e_basis}.
Section \ref{sec:results_disc} benchmarks NEO-MRCI results against NEO-FCI results for the \ce{HeHHe+} cation and presents results for three prototypical multicomponent systems: \ce{FHF-}, HCN, and HNC.
Section \ref{sec:conclusions} concludes and considers future directions.

\section{Theory and Computational Methods} \label{sec:theory_and_comp_meth}

The MR\emph{n}CI wavefunction is constructed via \emph{n} excitations into the virtual orbital space of an MCSCF reference, where $n = \text{S}, \text{SD}, ...$ for single, single and double, and so forth.\cite{Lischka2018}
The multiple virtual orbital spaces of a multicomponent CI wavefunction requires additional notation.
In this work we retain the convention that capital letters indicate excitation level but use subscripts to denote the included excitations.
Subscripts are omitted if all excitations of a type are included.
For example, $n = \text{S}\text{D}_\text{en}$ has all single electronic and nuclear excitations and double electronic-nuclear excitations, but no electronic-electronic double excitations.

This section provides a summary of these methods within the NEO framework.
Section \ref{subsec:mrci} introduces the NEO-CI wavefunction used in both NEO-MRCI and NEO-MCSCF, reviewing the formalism introduced in Ref.~\citenum{WebbIordanovSHS2002} and providing an overview of the algorithm.
Section \ref{subsec:orbital_opt} describes the orbital rotation step of the NEO-MCSCF method and outlines how both parts work in tandem to optimize electronic and nuclear molecular orbitals for a multiconfigurational reference space.\cite{SiegbahnAmlofHeibergRoos1981,WebbIordanovSHS2002,FajenBrorsen2021}
Section \ref{subsec:e_basis} discusses the revisions of  conventional electronic basis sets that are advantageous for these multicomponent methods.

\subsection{NEO-CI} \label{subsec:mrci}

The time-independent NEO-CI wavefunction is a linear combination of NEO configurations and can be expressed as
\begin{equation}
    \Psi^\text{NEO} (\mathbf{R}^\text{c}) = \sum_\mu c_\mu (\mathbf{R}^\text{c}) \psi_\mu^\text{NEO} (\mathbf{R}^\text{c}) ,
\end{equation}
where $\mathbf{R}^\text{c}$ are the coordinates of the classically treated nuclei.
In the following expressions, dependence on the classically treated nuclei is suppressed for clarity.
The NEO configurations, $\psi_\mu^\text{NEO}$, are formed from the product of individual electronic and nuclear Slater determinants:
\begin{equation}
    \psi_\mu^\text{NEO} = \lvert \Phi_{i(\mu)}^\text{e}\rangle \lvert \Phi_{I(\mu)}^\text{n} \rangle.
\end{equation}
$i(\mu)$ and $I(\mu)$ are the indices of the electronic and nuclear determinants, respectively, forming the $\mu$-th NEO configuration.
Dependence of the electronic (nuclear) determinants on the electronic (nuclear) coordinates is suppressed for clarity.

The NEO-CI adiabatic states are vibronic states that satisfy the eigenvalue equation
\begin{equation}
	\left[ \mathbf{H}^\text{NEO} - \mathbf{I}E^\text{NEO} \right] \mathbf{c} = \bf{0}, \label{eqn:neoseculareq}
\end{equation}
where $\mathbf{H}^\text{NEO}$ is the NEO Hamiltonian in the NEO configuration basis:
\begin{equation}
	H_{\mu\nu}^\text{NEO} = \langle \psi^\text{NEO}_\mu \rvert\hat{H}^\text{NEO}\lvert \psi^\text{NEO}_\nu \rangle \label{eqn:neomrci_ham}
\end{equation}
The NEO Hamiltonian operator, $\hat{H}^\text{NEO}$, for $N^\text{e}$ electrons, $N^\text{q}$ quantum nuclei, and $N^\text{c}$ classical nuclei is
\begin{equation}
	\hat{H}^\text{NEO} = \sum_{i=1}^{N^\text{e}}h^\text{e}(i) + \sum_{i=1}^{N^\text{e}}\sum_{j>i}^{N^\text{e}}\frac{1}{r_{ij}} + \sum_{I=1}^{N^\text{q}}h^\text{n}(I) + \sum_{I=1}^{N^\text{q}}\sum_{J>I}^{N^\text{q}}\frac{Z_I Z_J}{R_{IJ}} -\sum_{i=1}^{N^\text{e}}\sum_{I=1}^{N^\text{q}} \frac{Z_I}{r_{iI}} \label{eqn:neohamop}
\end{equation}
where $r_{ij}$ is the distance between electrons $i$ and $j$, $R_{IJ}$ is the distance between nucleus $I$ and nucleus $J$, and $Z_I$ is the charge of nucleus $I$. 
The electronic and nuclear one-particle terms, $h^\text{e}(i)$ and $h^\text{n}(I)$, are defined as
\begin{subequations}
    \begin{equation}
    h^\text{e}(i) = -\frac{1}{2}\nabla_i^2 - \sum_{J=1}^{N^\text{c}}\frac{Z_J}{r_{iJ}}
    \end{equation}
\text{and}
    \begin{equation}
    h^\text{n}(I) = -\frac{1}{2M_{I}}\nabla_I^2 + \sum_{J=1}^{N^\text{c}}\frac{Z_J Z_I}{R_{IJ}}
    \end{equation}
\end{subequations}
where $M_I$ is the mass of nucleus $I$.

The main effort in any CI calculation is computation of the matrix-vector product required by the Davidson algorithm\cite{Davidson1975} 
\begin{equation}
\mathbf{H} \mathbf{v} = \boldsymbol{\sigma}, \label{eqn:hv_sigma}
\end{equation}
where $\mathbf{H}$ is given in Eq.~(\ref{eqn:neomrci_ham}) and $\mathbf{v}$ is a trial vector.
Many algorithms exist for the efficient calculation of Eq.~(\ref{eqn:hv_sigma}).\cite{Shavitt1981,KnowlesHandy1984,FalesLevine2015,LishkaShepardMuller2020} 
Our NEO-CI implementation is a multicomponent extension of the string-based, determinant CI algorithm presented by Ivanic and Reudenberg.\cite{IvanicRuedenberg2001} 
This algorithm was chosen due to the ease with which it can be adapted for use in a NEO configuration basis and handle the truncated CI expansions necessary for MR\emph{n}CI and RASSCF wavefunctions.
In addition to the alpha and beta electron occupation strings that form electronic determinants, nuclear orbital occupation strings forming nuclear determinants are required.
In this implementation, all nuclei are assumed to have the same spin, so the maximum occupation of any nuclear orbital is 1.

The vector $\boldsymbol{\sigma}$ is the sum of ten contributions from the nonzero matrix elements of the NEO Hamiltonian: single and double replacements in the alpha electronic, beta electronic, and nuclear strings. 
Expressions for each of the ten contributions along with pseudocode outlining the evaluation of $\boldsymbol{\sigma}$ are presented in the Supporting Information (SI).

\subsection{Orbital Optimization: NEO-MCSCF} \label{subsec:orbital_opt}

The NEO multiconfigurational wavefunction energy can be expanded in a Taylor series parameterized by unitary matrices, $\mathbf{U}^\text{e}$ and $\mathbf{U}^\text{n}$, with elements
\begin{subequations}
    \begin{equation}
        \mathbf{U}^\text{e} = \text{exp}\left(\mathbf{X}^\text{e}\right)
    \end{equation}
    \begin{equation}
        \mathbf{U}^\text{n} = \text{exp}\left(\mathbf{X}^\text{n}\right)
    \end{equation}
\end{subequations}
where $\mathbf{X}^\text{e}$ and $\mathbf{X}^\text{n}$ are skew-symmetric matrices.
The transformation of the electronic ($\phi^\text{e}$) and nuclear ($\phi^\text{n}$) orbitals by $\mathbf{U}^\text{e}$ and $\mathbf{U}^\text{n}$, respectively, is expressed as
\begin{subequations}
    \begin{equation}
        \delta \phi^\text{e}_p = (\mathbf{U}^\text{e}\phi^\text{e})_p - \phi^\text{e}_p = \sum_q\left(X^\text{e}_{pq} + \frac{1}{2}\sum_r X^\text{e}_{pr}X^\text{e}_{rq} + \cdots\right)\phi^\text{e}_q
    \end{equation}
    \begin{equation}
        \delta \phi^\text{n}_P = (\mathbf{U}^\text{n}\phi^\text{n})_P - \phi^\text{n}_P = \sum_Q\left(X^\text{n}_{PQ} + \frac{1}{2}\sum_R X^\text{n}_{PR}X^\text{n}_{RQ} + \cdots\right)\phi^\text{n}_Q
    \end{equation}
\end{subequations}
$X^\text{e}_{pq}$ and $X^\text{n}_{PQ}$ are electronic and nuclear orbital rotation parameters, respectively.
Here lower-case indices correspond to electronic orbitals, and upper-case indices correspond to nuclear orbitals.
If
\[
\mathbf{X} = \begin{pmatrix} \mathbf{X}^\text{e} \\ \mathbf{X}^\text{n} \end{pmatrix},
\]
then to second order the NEO energy can be expressed as
\begin{equation}
	E^\text{NEO}(\mathbf{X}) = E^\text{NEO}(\mathbf{0}) + \mathbf{W}\mathbf{X} + \frac{1}{2}\mathbf{X}^\dagger \mathbf{A}\mathbf{X}, \label{eqn:neotaylor}
\end{equation}
where 
\begin{equation}
\mathbf{W} = \begin{pmatrix} \mathbf{W}^\text{e} \\ \mathbf{W}^\text{n} \end{pmatrix} \label{eqn:elec_and_nuc_gradient}
\end{equation}
and
\begin{equation}
\mathbf{A} = \begin{pmatrix} \mathbf{A}^\text{ee} & \mathbf{A}^\text{en} \\ \mathbf{A}^\text{ne} & \mathbf{A}^\text{nn} \end{pmatrix} \label{eqn:en_hessian}
\end{equation}
are the gradient and Hessian, respectively, of the energy with respect to electronic and nuclear orbital rotations.

The orbital rotational parameters $\mathbf{X}^\text{e}$ and $\mathbf{X}^\text{n}$ are coupled at second order via the off-diagonal blocks of the combined Hessian, $\mathbf{A}^\text{en} = \mathbf{A}^\text{ne}$.
In most multicomponent orbital optimization schemes, this coupling is ignored.\cite{PavosevicRousseauSHS2020,FajenBrorsen2021}
However, simultaneous optimizations have been shown to improve the convergence of multicomponent orbital optimizations and reduce computational cost.\cite{LiuChowWildmanFrischSHSLi2022}
Our implementation is a simultaneous optimization.

The NEO electronic and nuclear orbital gradients are
\begin{subequations}
\begin{equation}
	W_{pq}^\text{e} = 2\left(F_{pq}^{\text{NEO},\text{e}} - F_{qp}^{\text{NEO},\text{e}}\right) \label{eqn:elecOrbitalGradient}
\end{equation}
and
\begin{equation}
    W_{PQ}^\text{n} = 2\left(F_{PQ}^{\text{NEO},\text{n}} - F_{QP}^{\text{NEO},\text{n}}\right)\label{eqn:nucOrbitalGradient}
\end{equation}
\end{subequations}
where $\mathbf{F}^{\text{NEO},\text{e}}$ and $\mathbf{F}^{\text{NEO},\text{n}}$ are the NEO generalized electronic and nuclear Fock matrices, respectively, with elements
\begin{subequations}
\begin{equation}
	F_{pq}^{\text{NEO},\text{e}} = \sum_r D_{pr}^\text{e} h^\text{e}_{qr} + 2 \sum_{rst} P_{prst}^\text{e}\left(qr \vert st\right) - \sum_r \sum_{PQ} P_{prPQ}^\text{en} \left(qr \vert PQ\right) \label{eqn:elecNeoFockpq}
\end{equation}
\begin{equation}
F_{PQ}^{\text{NEO},\text{n}} = \sum_R D_{PR}^\text{n} h^\text{n}_{QR} + \sum_{RST} P_{PRST}^\text{n}\left(QR \vert ST\right) - \sum_R \sum_{pq} P_{pqPR}^\text{en} \left(pq \vert QR\right) \label{eqn:nucNeoFockpq}
\end{equation}
\end{subequations}
Here $\mathbf{D}^\text{e}$ and $\mathbf{P}^\text{e}$ are the electronic one- and two-particle reduced density matrices, $\mathbf{D}^\text{n}$ and $\mathbf{P}^\text{n}$ are the nuclear one- and two-particle reduced density matrices, and $\mathbf{P}^\text{en}$ is the mixed electronic-nuclear two-particle reduced density matrix.
Moreover, $\left(qr \vert st\right)$ denote two-electron integrals, $\left(QR \vert ST \right)$ denote two-nucleus integrals, and $\left(qr \vert PQ\right)$ denote mixed-particle, electron-nucleus integrals in chemist's notation.

The NEO orbital Hessian is composed of electronic, nuclear, and nuclear-electronic blocks:
\begin{subequations}
    \begin{equation}
	   A_{pq,rs}^\text{ee} = (1 - \tau_{pq})(1 - \tau_{rs})a^\text{e}_{pqrs},\label{eqn:elecOrbitalHessian}
    \end{equation}
    \begin{equation}
        A_{PQ,RS}^\text{nn} = (1 - \tau_{PQ})(1 - \tau_{RS})a^\text{n}_{PQRS},\label{eqn:nucOrbitalHessian}
    \end{equation}
    \begin{equation}
        A_{pq,PQ}^\text{en} = (1 - \tau_{pq})(1 - \tau_{PQ})a^\text{en}_{pqPQ}. \label{eqn:mixedOrbitalHessian}
    \end{equation}
\end{subequations}
where
\begin{subequations}
    \begin{equation}
        \begin{split}
            a^\text{e}_{pqrs} = 2D^\text{e}_{pr}h^\text{e}_{qs} + 4\sum_{tu}\left[ P^\text{e}_{rptu} \pint{qs}{tu} + \left( P^\text{e}_{rutp} + P^\text{e}_{rupt}\right)\pint{qt}{su}\right]\\ + 2\sum_{PQ}P^\text{en}_{qrPQ}\pint{sp}{PQ} + \delta_{qr}\left(F^{\text{NEO},\text{e}}_{ps} + F^{\text{NEO},\text{e}}_{sp}\right)
        \end{split}
    \end{equation}
    \begin{equation}
        \begin{split}
            a^\text{n}_{PQRS} = 2D^\text{n}_{PR}h^\text{n}_{QS} + 4\sum_{TU}\left[ P^\text{n}_{RPTU} \pint{QS}{TU} + \left( P^\text{n}_{RUTP} + P^\text{n}_{RUPT}\right)\pint{QT}{SU}\right]\\ + 2\sum_{pq}P^\text{en}_{pq QR}\pint{pq}{SP} + \delta_{QR}\left(F^{\text{NEO},\text{n}}_{PS} + F^{\text{NEO},\text{n}}_{SP}\right)
        \end{split}
    \end{equation}
    \begin{equation}
        \begin{split}
            a^\text{en}_{pqPQ} = -2\sum_r\sum_R \left[P^\text{en}_{pr PR}\pint{qr}{QR} - P^\text{en}_{pr QR}\pint{qr}{PR}\right]
        \end{split}
    \end{equation}
\end{subequations}
are general expressions for each subblock of $\mathbf{A}$.
$\tau_{pq}$ is the permutation of orbital indices $p$ and $q$.
Explicit expressions for $\mathbf{F}^{\text{NEO},\text{e}}$, $\mathbf{F}^{\text{NEO},\text{n}}$, $\mathbf{D}^\text{e}$, $\mathbf{D}^\text{n}$, $\mathbf{P}^\text{e}$, $\mathbf{P}^\text{n}$,$\mathbf{P}^\text{en}$, $\mathbf{A}^\text{ee}$, $\mathbf{A}^\text{nn}$ and $\mathbf{A}^\text{en}$ are presented in the SI.

The orbital optimization seeks the minimum of Eq.~(\ref{eqn:neotaylor}), which is determined by solving the Newton-Raphson equation: 
\begin{equation}
	\mathbf{0} = \mathbf{W} + \mathbf{A}\mathbf{X}. \label{eqn:NewtonRaphson}
\end{equation}
Eq.~(\ref{eqn:NewtonRaphson}) is solved iteratively via the augmented Hessian method\cite{Yarkony1981, KreplinKnowlesWerner2019,HelmichParis2021} or gradient-descent minimization, which requires only the diagonal elements of Eq.~(\ref{eqn:en_hessian}).
Details about the augmented Hessian method are provided in the SI.

The NEO-MCSCF method that we implemented is a decoupled two-step process,\cite{SiegbahnAmlofHeibergRoos1981,KreplinKnowlesWerner2019} neglecting the coupling between orbital rotations and CI expansion coefficients. Figure \ref{fig:neoMCalgo} outlines this algorithm.
The density matrices $\mathbf{D}^\text{e}$, $\mathbf{D}^\text{n}$, $\mathbf{P}^\text{e}$, $\mathbf{P}^\text{n}$, and $\mathbf{P}^\text{en}$ are computed from the NEO-CI wavefunction each macroiteration.
In the state-averaged case, the weighted sums of the density matrices for each state are used. This suite of multireference NEO methods is implemented in a developer version of Q-Chem.\cite{QChem5Cite}

\begin{figure}
\centering
\textbf{NEO-MCSCF Algorithm}\par\medskip
\begin{algorithmic}[ht]
\WHILE{$\Delta E^\text{NEO} > t_E$}
	\STATE $\text{Solve }\left[\mathbf{H}^\text{NEO} - \mathbf{I}E^\text{NEO}\right]\mathbf{c} = \mathbf{0}$
	\STATE $\text{Compute } \mathbf{D}^\text{e}, \mathbf{D}^\text{n}, \mathbf{P}^\text{e}, \mathbf{P}^\text{n}, \mathbf{P}^\text{en}, E^\text{NEO}$
    \WHILE{$\|\mathbf{W}\| > t_W$}
	\STATE $\mathbf{X} = -\mathbf{A}^{-1}\mathbf{W}$
	\STATE $\text{Update } \mathbf{C}^\text{e}, \mathbf{C}^\text{n}$
    \ENDWHILE
\ENDWHILE
\end{algorithmic}
\caption{   NEO-MCSCF algorithm implemented in this work. 
            $t_E$ and $t_W$ are the convergence tolerances for the change in NEO-MCSCF energy and the norm of the orbital gradients, respectively. 
            $\mathbf{C}^\text{e}$ and $\mathbf{C}^\text{n}$ are the electronic and nuclear molecular orbital coefficients, and ${\bf c}$ are the coefficients of the NEO configurations in the CI expansion.}
\label{fig:neoMCalgo}
\end{figure}

\subsection{Electronic Basis Set Considerations} \label{subsec:e_basis}

In multicomponent calculations, relatively large electronic basis sets, e.g, cc-pV5Z,\cite{PetersonWoonDunning1994} are often centered on the quantized nuclei.\cite{CulpittYangPavosevicTaoSHS2019,Brorsen2020,YuPavosevicSHS2020,FajenBrorsen2021}
Large electronic basis sets are required to describe the electronic structure of a delocalized nucleus, which is no longer a static point charge.
However, adding more basis functions to existing electronic basis sets \cite{SamsonovaTuckerAlaalBrorsen2023} does not significantly improve the agreement with numerically exact grid-based benchmarks.

Standard electronic basis sets are designed within the Born-Oppenheimer approximation.
The tightest s-function describes the core 1s orbital, with a radial component decaying quickly from the static point charge nucleus.
This tight s-function is designed to replicate Kato's cusp condition:\cite{Kato1957} several Gaussian functions are contracted to model a cusp at the nuclear center.
This cusp represents the physics of the electronic density in the field of a fixed, classical point-charge nucleus, but it does not represent the physics of the electronic density surrounding a quantized, delocalized nucleus.\cite{BochevarovValeevSherrill2004}
As there are no static point-charge positions for the quantized nuclei in NEO calculations, {\color{revisions}there should be no single cusp. 
Because the electronic and nuclear densities coincide at many points, treating one point differently from all other points by introducing a single cusp at the proton basis function center is an unbalanced approach that can lead to unphysical behavior.}
To alleviate this non-physical behavior of conventional electronic basis sets in NEO calculations, we propose removal of the tight Gaussian primitives that compose the contracted core 1s basis function, leaving only the most diffuse Gaussian of the contraction to represent this 1s orbital.
In this work, this modification of conventional electronic basis sets will be denoted with an asterisk (*), e.g., the modified def2-QZVP\cite{Weigend2006} basis set will be denoted def2-QZVP*.

Additionally, electronic basis sets specifically designed for multicomponent calculations are expected to be uniformly more diffuse than their conventional electronic analogs, as the delocalized nucleus is a charge distribution instead of a static point charge.
In relativistic theory, basis sets such as the Dyall family\cite{Dyall2016} are optimized to describe the relativistic electronic structure of a nuclear charge distribution.\cite{DyallFaegri1993,VisscherDyall1997,DyallFaegri2007,SunStetinaZhangLi2021}
However, as the nuclear position is still static, {\color{revisions}even these electronic basis sets are not suitable for describing the quantum nuclei in multicomponent calculations. 
Although the Dyall basis sets may offer an improved description of predominantly spherically symmetric ground state protonic densities, they do not correctly describe asymmetric ground state protonic densities or the excited vibronic states, where the proton vibrational wavefunctions have nodes and often are even more asymmetric.}
Because the development of new electronic basis sets is beyond the scope of this work, we apply a uniform multiplicative factor to the exponents of the conventional electronic basis sets.
The scaling factor, $\gamma$, is chosen to minimize the root mean square error (RMSE) of the NEO single-reference (NEO-HF molecular orbitals) single and double CI with electron-electron-nucleus triple excitations (SDT$_\text{een}$CI) ground state protonic density compared with numerically exact FGH results.
For closed-shell systems that are predominantly single configurational, the NEO-SDT$_\text{een}$CI ground state result will be similar to the NEO full CI (NEO-FCI) result.

\begin{figure}[htp]
\centering
\includegraphics[width=4in]{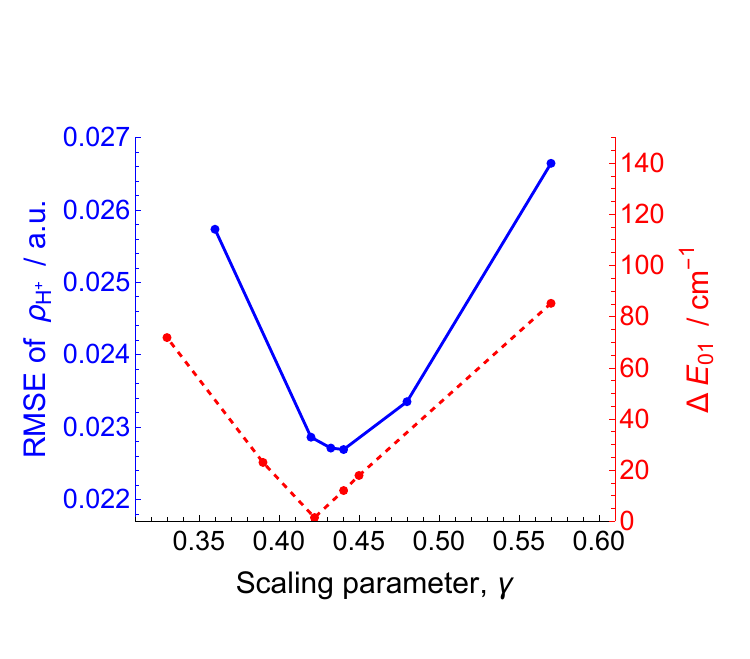}
\caption{For the molecule \ce{HeHHe+}, RMSE of the ground state protonic density computed with NEO-SDT$_\text{een}$CI and the absolute error of the first excited vibronic state excitation energy computed with NEO-FCI compared to the FGH benchmark as a function of the electronic basis set scaling factor $\gamma$.
The electronic and protonic basis sets for H are $\gamma$-def2-QZVP* and 8s8p8d8f, respectively.
The electronic basis set for He is 6-31G.
The He-He distance is 1.8 \si{\angstrom}. The reference FGH calculation used FCI and the 6-31G electronic basis set on all atoms.}
\label{fig:ebasis_scaling_def2qzvp}
\end{figure}

We have found that the accuracy of the ground state protonic density is correlated with the accuracy of the vibronic state excitation energies when the electronic basis set associated with the quantum proton is optimized in this manner.
Figure \ref{fig:ebasis_scaling_def2qzvp} shows the root-mean-square error (RMSE) of the ground state protonic density computed with NEO-SDT$_\text{een}$CI and the absolute error of the NEO-FCI first excited state excitation energy for each value of $\gamma$ for \ce{HeHHe+}. 
The He nuclei are fixed at a separation of 1.8 \si{\angstrom}, and the H nucleus is treated quantum mechanically with its electronic and protonic basis function centers fixed at the midpoint.
The optimal scaling parameter for the ground state protonic density also approximately minimizes the error of the first excited vibronic state excitation energy.
The scaled basis sets will be denoted by the prefix $\gamma$-.
In this work, the modified and scaled def2-QZVP electronic basis set is denoted $\gamma$-def2-QZVP*.
Exponent parameters for the revised electronic basis sets used in this work are presented in the SI.

{\color{revisions}
We emphasize that these electronic basis sets are specifically designed to describe protonic densities and proton vibrational excitation energies in multicomponent quantum chemistry calculations. The goal is to reproduce the results from a numerically exact treatment, in which the electronic basis function centers are moved with the proton on a grid, effectively providing a complete electronic basis set associated with the proton. Multicomponent orbital methods often represent the electronic basis set associated with the quantum proton by a single basis function center, which is usually shared by the proton basis set. In this case, we need to design the electronic basis set in a manner that circumvents the imbalance caused by the electronic basis functions being centered at a single point.
From a physical perspective, this imbalance in the treatment of points spanned by the nuclear wavefunction causes the electronic and protonic densities to localize where the electronic wavefunction flexibility is greatest for such localizations, resulting in ground state protonic densities that are too localized.
This imbalance causes the descriptions of excited vibronic states to be even worse because the fundamental excited protonic vibrational wavefunctions have nodes at or near the position of the protonic basis function center. Thus, for excited vibronic states, the electronic wavefunction has the most flexibility where the nuclear density is zero or nearly zero.
This unequal treatment of ground and excited proton vibrational states leads to significantly overestimated excitation energies (Table S3).

Our solution is to omit the tight core electronic basis functions and to scale the other basis functions in a way that provides accurate ground state protonic densities compared to the numerically exact grid-based reference. By reproducing the protonic
density of the ground vibronic state, the revised electronic basis set is approximating
the electronic environment that would be obtained with a complete basis set. As shown above, accurate ground state protonic densities are correlated with accurate vibronic state excitation energies. We acknowledge that the variationally obtained ground vibronic state energy will be lower when additional, more compact electronic basis functions are added to the basis set, but the physical properties such as protonic densities and vibronic state excitation energies will become much worse when this variational flexibility is permitted. We have found that restricting the electronic basis set to maintain a balanced description of delocalized protonic densities in both the ground and excited vibronic states is essential for excited-state multicomponent wavefunction methods using a single basis function center for each quantum proton. Note that we do not expect these revised electronic basis sets to provide optimal results for conventional electronic structure calculations with point-charge nuclei because they are designed expressly for the case of delocalized protonic densities.
}

\subsection{The Fourier grid Hamiltonian Benchmark} \label{subsec:FGH}
The FGH calculations used as the benchmark for protonic densities and proton vibrational excitation energies in this work are numerically exact at the specified level of electronic structure for electronically adiabatic systems.
In these FGH calculations, single-point energies are computed as the hydrogen nucleus is moved on a three-dimensional (3D) grid with all other nuclei fixed.
Subsequently, a 3D Schr\"odinger equation is solved for the proton moving on this 3D PES, providing the proton vibrational wavefunctions and energy levels.
The electronic basis functions associated with the proton move with the proton in the generation of the PES to ensure a fully converged electronic basis set.
If the electronic basis functions of the proton were to be fixed at a single point,\cite{NareshBrorsen2021} such as the minimum on the 3D PES, the electronic basis set would be inadequate to describe most geometries associated with the grid.
As a result, the proton vibrational excitation energies would be much too high. Specifically, these excitation energies would be significantly higher than those computed with a harmonic, normal-mode analysis or measured experimentally.
Thus, the original, numerically exact FGH method\cite{MarstonBalintKurti1989,WebbSHS2000} is used in this work.

\section{Results and Discussion} \label{sec:results_disc}
This section presents applications of the NEO-MRCI approach to four different molecular systems. 
As discussed in Sec.~\ref{subsec:FGH}, benchmarks are provided by FGH calculations, which are numerically exact for electronically adiabatic systems.\cite{MarstonBalintKurti1989, WebbSHS2000} 
Note that the proton vibrational excitation energies cannot be compared directly to experimental spectra because only the proton is treated quantum mechanically, and the classical nuclei are fixed. 
The motion of the classical nuclei and the coupling between the classical and quantum nuclei can be included with nonadiabatic dynamics methods such as Ehrenfest or surface hopping dynamics.\cite{CrespoBarbatti2018} 
Such nonadiabatic dynamics methods have been implemented within the NEO framework in conjunction with real-time NEO-TDDFT\cite{ZhaoWildmanTaoSchneiderSHSLi2020,ZhaoWildmanPavosevicTullySHSLi2021} and NEO-MSDFT\cite{YuRoySHS2022,DickinsonSHS2024} but not yet with multireference wavefunction methods.

\subsection{\ce{HeHHe+}} \label{subsec:hehhe}
\subsubsection{Full CI Benchmark}
The \ce{HeHHe+} system has been used as a benchmark for numerous multicomponent methods.\cite{WebbIordanovSHS2002,FeldmannMuoloBaiardiReiher2022,GarnerUpadhyayLiSHS2024}
Due to its small size, novel multicomponent methods can be compared with FCI results, which represent the limit for given electronic and nuclear basis sets.
This comparison is important for assessing the accuracy of truncated CI methods.
At best, a method can return the FCI result.
The FCI method will also serve to verify our revised electronic basis sets. 
For these calculations, the 6-31G\cite{DitchfieldHehrePople1971} electronic basis set was used for the He atoms, which were separated by 1.8 \AA. 
{\color{revisions}Note that this relatively small electronic basis set was only used for the \ce{HeHHe+} system to
allow NEO-FCI calculations. 
Larger electronic basis sets on the classical nuclei were used for the
other molecular systems studied.}
The procedure described in Sec.~\ref{subsec:e_basis} was used to determine the $\gamma$ value of 0.4404 in the electronic basis set, denoted 0.4404-def2-QZVP*, for the quantum proton. 
The even-tempered 8s8p8d8f protonic basis set\cite{BardoRuedenberg1974,YangBrorsenCulpittPakSHS2017} was used.
{\color{revisions} Larger electronic and protonic basis sets were used to describe the overtone vibrations for \ce{HeHHe+} below.}

Table \ref{tbl:HeHHe_results} reports the ground state energy and the excitation energy of the first excited vibronic state for \ce{HeHHe+} computed with the NEO-SA-MCSCF, NEO-MR-SD$_\text{en}$CI, and NEO-MR-SDT$_\text{een}$CI methods. 
The first excited vibronic state of \ce{HeHHe+} corresponds to the proton bend vibrational mode, as shown in Figure \ref{fig:hehhe_modes}a.
For a modestly sized electronic active space of 4 electrons in 6 orbitals (225 electronic determinants), the NEO-SA-MCSCF method is able to obtain only qualitative agreement with the reference FGH result.
However, NEO-MR-SD$_\text{en}$CI and NEO-MR-SDT$_\text{een}$CI vibronic excitation energies are in good agreement with the reference FGH result, reporting errors of +34 \si{\wn} and +15 \si{\wn}, respectively.
More importantly, the agreement with the NEO-FCI result, which represents the limit for the basis sets used to describe this system, is excellent.
The NEO-MR-SD$_\text{en}$CI method does not recover the correlation energy of the NEO-FCI method, but it does provide vibronic excitation energies nearing NEO-MR-SDT$_\text{een}$CI accuracy.
For some systems, the NEO-MR-SDT$_\text{een}$CI method will be prohibitively expensive. 
In these cases, the NEO-MR-SD$_\text{en}$CI method can be used to compute accurate vibronic excitation energies at a fraction of the cost.
The nuclear-electronic all-particle (NEAP) DMRG included in Table \ref{tbl:HeHHe_results} used conventional electronic basis sets, and the agreement with the FGH reference would likely improve significantly if the electronic basis sets from Sec.~\ref{subsec:e_basis} were used.

\begin{figure}[htp]
\centering
\includegraphics[width=.99\linewidth]{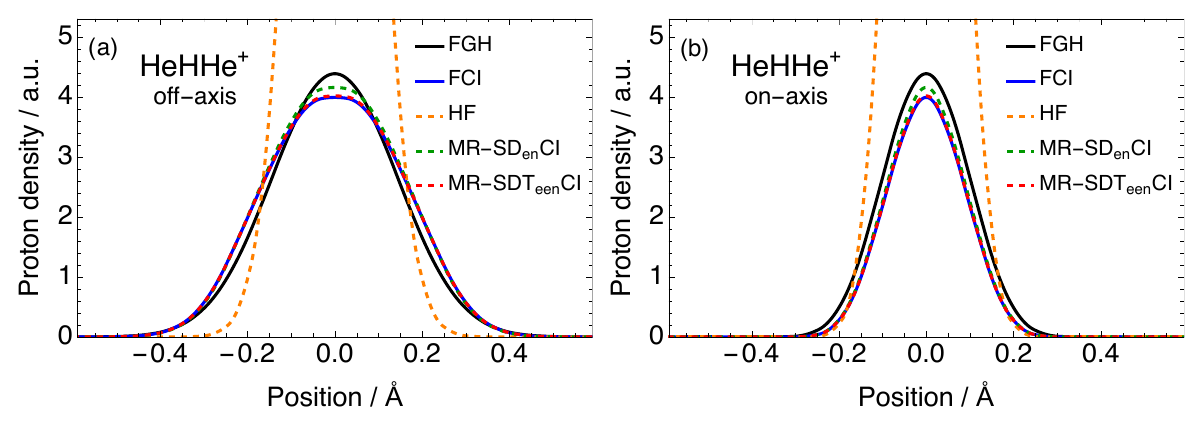}
\caption{Ground state protonic density (a) off-axis and (b) on-axis for \ce{HeHHe+} computed with the NEO-MR-SD$_\text{en}$CI and NEO-MR-SDT$_\text{een}$CI methods compared to NEO-FCI, NEO-HF, and reference FGH results.
The distance between the fixed He atoms is 1.8 \si{\angstrom} with the midpoint at the origin.
The electronic basis set for He is 6-31G.
The electronic and protonic basis sets for H are 0.4404-def2-QZVP* and 8s8p8d8f, respectively, and are placed at the origin.
The electronic active space is 6 orbitals and 4 electrons, and the protonic active space is 87 orbitals and 1 proton.
The reference FGH calculation uses FCI and the 6-31G electronic basis set for all atoms.
The grid spans the range of -0.8203 to 0.8750 \si{\angstrom} around the origin.
The NEO-HF maximum protonic density is 21 a.u.}
\label{fig:hehhe_gs_density_compare}
\end{figure}

The NEO-MRCI method also describes proton delocalization accurately.
In Figure \ref{fig:hehhe_gs_density_compare}, the ground state protonic density computed at the NEO-MR-SD$_\text{en}$CI and NEO-MR-SDT$_\text{een}$CI levels is compared to the NEO-FCI protonic density and the reference FGH benchmark.
The agreement is excellent.
Linear response multicomponent methods such as NEO-TDDFT have obtained accurate vibronic excitation energies but do so with a ground state protonic density that is too localized,\cite{YuPavosevicSHS2020} suggesting a cancellation of errors when computing excitation energies.
The NEO-MRCI method agrees quantitatively with the NEO-FCI and FGH benchmarks for both the vibronic excitation energies and the ground state protonic densities.
Thus, the NEO-MRCI wavefunction provides a more accurate description of the physics of the multicomponent system.

\subsubsection{Overtone Vibrations}

\begin{figure}[htp]

    \includegraphics[trim={0, 0, 0, 0},clip,width=3.2in]{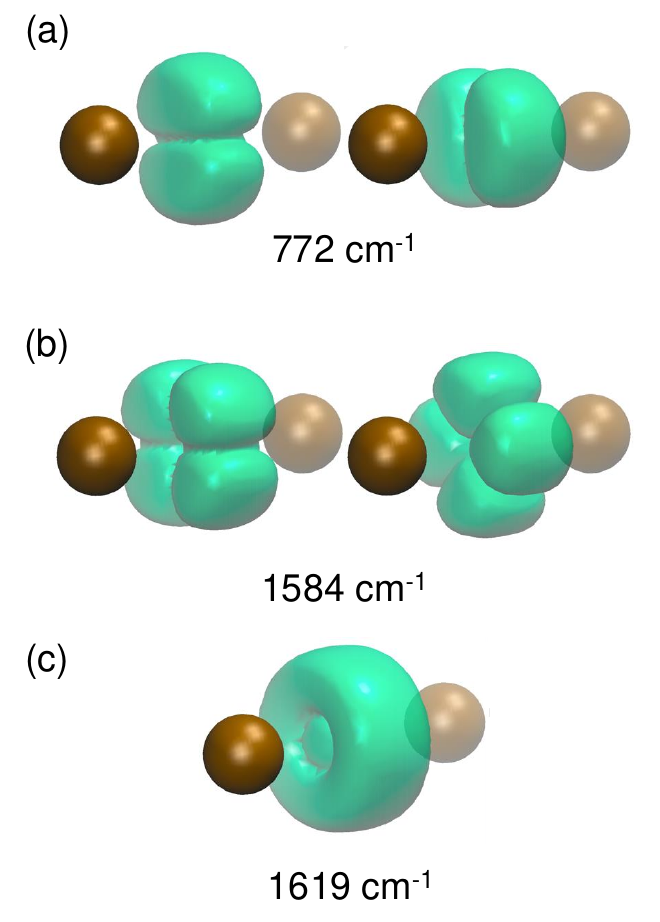}
\caption{Proton vibrational bend modes for \ce{HeHHe+} computed using the FGH method in conjunction with conventional electronic CCSD. (a) fundamental modes, (b) overtone modes, and (c) combination mode.}
\label{fig:hehhe_modes}

\end{figure}

The eigenstates of the NEO-CI Hamiltonian include overtone vibrations.
Describing such states is challenging because the protonic charge is delocalized over a greater volume than for the ground and fundamental vibrational states.
In previous work using NEO equation-of-motion coupled cluster methods, the excitation energies associated with overtones and combination modes are greatly overestimated.\cite{PavosevicTaoCulpittZhaoLiSHS2020}
The overtone vibrations of \ce{HeHHe+} computed with NEO-MR-SDT$_\text{een}$CI are reported in Table \ref{tbl:HeHHe_overtone_results} and shown in Figure \ref{fig:hehhe_modes}b-c.
We adopt the convention that a vibrational state is denoted by $(\nu_1, \nu_2, \nu_3)$, where $\nu_1$ and $\nu_2$ are the $x$- and $y$-axis bends and $\nu_3$ is the fundamental stretch.\cite{PavosevicTaoCulpittZhaoLiSHS2020} 
To describe the higher-order vibrations, the even-tempered protonic basis set is extended to include $g$-type functions for a total of 116 protonic basis functions.
The electronic basis set for He was aug-cc-pVTZ, which is large enough to describe the electronic structure of the cation.
The molecular orbitals were optimized via NEO-SA-MCSCF in a reference space of 4 electrons in 8 orbitals and 1 proton in 116 orbitals, averaging the lowest 7 states.
The NEO-MR-SDT$_\text{een}$CI expansion for this reference space is 52 751 232 NEO configurations.
The $(2,0,0)$ and $(0,2,0)$ overtone vibrations and the $(1, 1, 0)$ combination vibration are 70 \si{\wn} and 165 \si{\wn} above the reference, respectively. The fundamental stretch mode is higher than these modes and is the sixth excited vibronic state. Its frequency is qualitatively but not quantitatively accurate, with a computed value of 2371 \si{\wn} compared to the FGH benchmark value of 1818 \si{\wn}.

\begin{table}
\let\TPToverlap=\TPTrlap
\centering
\caption{NEO-SA-MCSCF, NEO-MR-SD$_\text{en}$CI, and NEO-MR-SDT$_\text{een}$CI Ground State Energy and Excitation Energy  of Fundamental Bend Modes for \ce{HeHHe+}\tnote{\emph{a}}}
\label{tbl:HeHHe_results}
\begin{threeparttable}
\begin{tabularx}{1\textwidth}{l Z Y Y}
    \toprule
    \toprule

    Method & Configurations & $E_0$ (Ha)  & $E_1 - E_0$ (\si{\wn}) \\
    \midrule
    NEO-SA-MCSCF\tnote{\emph{b}}            &     19 575 & -5.79413963  & 1271\\
    NEO-MR-SD$_\text{en}$CI          &    458 055 & -5.82396888  & 805\\
    NEO-MR-SDT$_\text{een}$CI        &  3 900 123 & -5.83799110  & 786\\
    \vspace{1\baselineskip}\\
    NEO-FCI                 & 27 380 727 & -5.83808345  & 783\\
    \vspace{1\baselineskip}\\
    FGH Reference\tnote{\emph{c}}      &  &  &771\\
    NEAP-DMRG\tnote{\emph{d}}      &  &  &1145\\
    \bottomrule 
\end{tabularx}
\begin{tablenotes}\linespread{1}\footnotesize
    \item[\emph{a}]    The distance between the two He atoms is 1.8 \si{\angstrom}. The electronic basis set for He is 6-31G.
                The electronic and protonic basis sets for H are 0.4404-def2-QZVP* and 8s8p8d8f, respectively.
                The electronic active space is 6 orbitals and 4 electrons.
                The protonic active space is 87 orbitals and 1 proton. 
                The first excited vibronic state corresponds to the proton bend mode.
    \item[\emph{b}]    The NEO-SA-MCSCF averages over the lowest three vibronic states.
    \item[\emph{c}]    The FGH calculation uses FCI and the 6-31G electronic basis set on all atoms.
                The grid spans the range of -0.8203 to 0.8750 \si{\angstrom} around the midpoint.
    \item[\emph{d}]    Result from Ref.~\citenum{FeldmannMuoloBaiardiReiher2022} using a conventional electronic basis set. 
    See Table S1 for a direct comparison of methods using the same electronic basis set.
\end{tablenotes}
\end{threeparttable}
\end{table}

\begin{table}
\let\TPToverlap=\TPTrlap
\centering
\caption{Proton Bend Mode Excitation Energies for \ce{HeHHe+} Computed with NEO-MR-SDT$_\text{een}$CI Compared to FGH Benchmark\tnote{\emph{a}}}
\label{tbl:HeHHe_overtone_results}
\begin{threeparttable}
\begin{tabularx}{1\textwidth}{W Y Y}
    \toprule
    \toprule

    Vibrational State & FGH\tnote{\emph{b}} & NEO-MR-SDT$_\text{een}$CI \\
    \midrule
    $(1, 0, 0)$, $(0, 1, 0)$  &  772  & 723 \\
    $(2, 0, 0)$, $(0, 2, 0)$  & 1584  & 1654\\
    $(1, 1, 0)$               & 1619  & 1784\\
    \bottomrule 
\end{tabularx}
\begin{tablenotes}\linespread{1}\footnotesize
    \item[\emph{a}]    The energies are given in \si{\wn}. The distance between the two He atoms is 1.8 \si{\angstrom}. 
                The electronic basis set for He is aug-cc-pVTZ.
                The electronic and protonic basis sets for H are 0.4404-def2-QZVP* and 8s8p8d8f8g, respectively.
                The electronic active space is 4 electrons in 8 orbitals.
                The protonic active space is 116 orbitals and 1 proton. 
                The orbitals are optimized with NEO-SA-MCSCF averaging over the lowest seven vibronic states.
    \item[\emph{b}]    The reference is from an FGH calculation using CCSD and the aug-cc-pVTZ electronic basis set on all atoms.
                The grid spans the range of -0.8203 to 0.8750 \si{\angstrom} around the midpoint.
\end{tablenotes}
\end{threeparttable}
\end{table}

\subsection{\ce{FHF-}}

The electronic structure of the \ce{FHF-} anion requires diffuse functions. \cite{HirataYagiAjithYamazakiHirao2008,DunningXu2021}
The electronic basis set for the flourine atoms is aug-cc-pVTZ.\cite{KendallDunningHarrison1992}
For the quantized proton, the electronic basis set is 0.4742-def2-QZVP*, where the $\gamma$ value of 0.4742 was calculated via the procedure outlined in Section \ref{subsec:e_basis}.
We note that $\gamma=0.4742$ is close to the value used for the \ce{HeHHe+} cation ($\gamma=0.4404$) in Section \ref{subsec:hehhe}.
Both protons are internal, and therefore the electronic environment is similar enough that the electronic basis set for one system is likely suitable for the other.
{\color{revisions} Table S5 and Figures S9 and S10 show the transferability of the revised electronic basis set for internal hydrogen nuclei.}
The protonic basis set is the even-tempered 8s8p8d8f basis set.

\begin{figure}[htp]
\includegraphics[trim={0, 0, 0, 0},clip,width=3.2in]{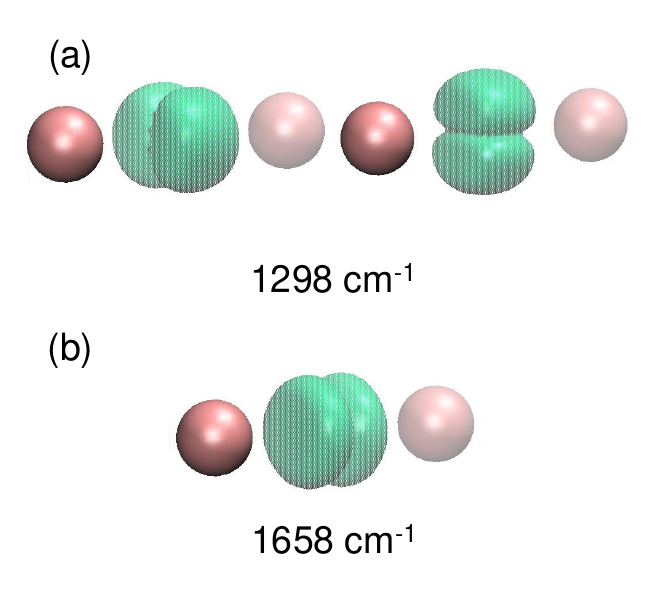}
\caption{Proton vibrational modes for \ce{FHF-} computed using the FGH method in conjunction with conventional electronic CCSD. (a) degenerate fundamental bend modes and (b) fundamental stretch mode.}
\label{fig:fhf_modes}
\end{figure}

The orbitals used in the NEO-MR-SD$_\text{en}$CI calculation are from a NEO-SA-MCSCF calculation, where the lowest four states are weighted equally.
These states represent the ground vibrational state and the fundamental degenerate proton bend modes and the proton stretch mode (Figure \ref{fig:fhf_modes}).
The electronic active space is composed of two $\sigma$ bonding orbitals and two $\sigma^*$ non-bonding orbitals to describe the motion of the proton along the molecular axis.
The protonic active space is composed of 20 orbitals.
The final NEO-MR-SD$_\text{en}$CI wavefunction is constructed via single excitations into the virtual space of both electronic and protonic reference wavefunctions, resulting in an expansion of 6 459 228 NEO configurations.
The fluorine 1s orbitals remain doubly occupied in all configurations.

The fundamental proton bend and stretch mode excitation energies computed with NEO-MR-SD$_\text{en}$CI for \ce{FHF-} are reported in Table \ref{tbl:fhf_FundFreq_results}.
For the proton bend mode, the agreement with the FGH benchmark  using conventional electronic CCSD to compute the 3D PES is reasonable.
However, the energy of the stretch mode is overestimated by 413 \si{\wn}.
This large error can be ascribed to possible differences in the electronic description between CCSD and MR-SCI.
The CCSD/aug-cc-pVTZ equilibrium F---F distance is $R_\text{FF} = 2.271$ \si{\angstrom}, whereas the MR-SCI/aug-cc-pVTZ equilibrium F---F distance is $R_\text{FF} = 2.320$ \si{\angstrom}.
The excitation energy of the stretch mode computed with NEO-MR-SD$_\text{en}$CI at the longer distance of $R_\text{FF} = 2.320$ \si{\angstrom} is overestimated by only 126 \si{\wn}.
Although the NEO-MR-SD$_\text{en}$CI excitation energies are overestimated for this system, they are the best results obtained with a multicomponent wavefunction method to date.
The multicomponent HCI results\cite{NareshBrorsen2021} included in Table \ref{tbl:fhf_FundFreq_results} used conventional electronic basis sets, and the agreement with the FGH reference would likely improve significantly if the electronic basis sets from Sec.~\ref{subsec:e_basis} were used.

\begin{figure}[htp]
\centering
\includegraphics[width=.99\linewidth]{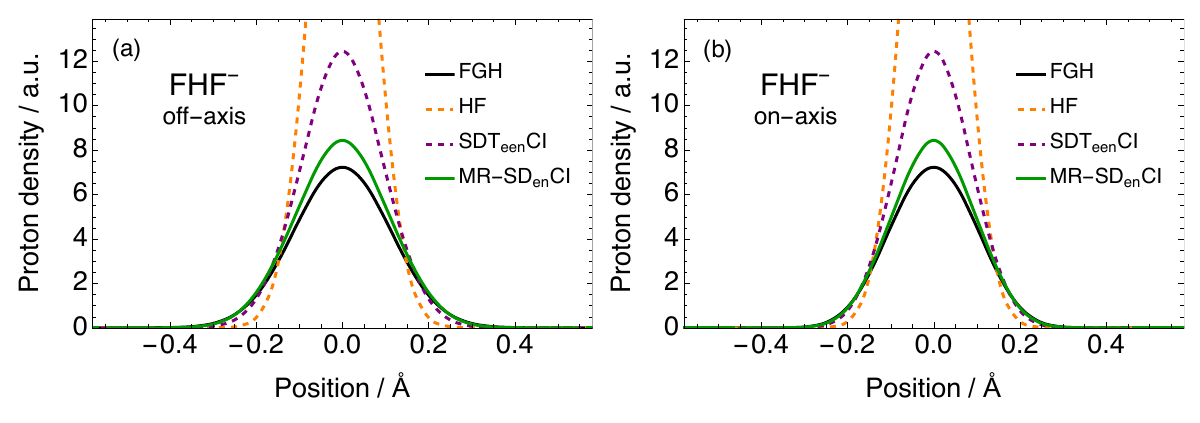}
\caption{Ground state protonic density (a) off-axis and (b) on-axis for \ce{FHF-} computed with the NEO-MR-SD$_\text{en}$CI, NEO-SDT$_\text{een}$CI, and NEO-HF methods compared to the reference FGH results. 
The distance between the fixed F atoms is 2.271  \si{\angstrom} with the midpoint at the origin. 
The electronic basis set for F is aug-cc-pVTZ.
The electronic and protonic basis sets for H are $0.4742$-def2-QZVP* and 8s8p8d8f, respectively, and are placed at the origin.
The electronic active space is 4 orbitals and 4 electrons, and 
the protonic active space is 20 orbitals and 1 proton.
The reference FGH calculation uses CCSD and the aug-cc-pVTZ electronic basis set for all atoms.
The grid spans the range of -0.7275 to 0.7760 \si{\angstrom} around the origin.
The NEO-HF maximum protonic density is 29 a.u.}
\label{fig:fhf_gs_density_compare}
\end{figure}

In addition to accurate vibronic state excitation energies, the protonic density is also well described by the NEO-MR-SD$_\text{en}$CI method for this system.
Figure \ref{fig:fhf_gs_density_compare} shows that the agreement between the NEO-MR-SD$_\text{en}$CI method and the reference FGH method is excellent for the ground state protonic density. For comparison, the NEO-HF protonic density has a much higher maximum value of 29 a.u.
The NEO-MR-SD$_\text{en}$CI protonic density of the stretch mode is compared with the FGH result in Figure S6.
The over-localization of the protonic density for this mode comports with the overestimation of the excited state energy shown in Table \ref{tbl:fhf_FundFreq_results}.

\begin{table}
\let\TPToverlap=\TPTrlap
\centering
\caption{Fundamental Vibrational Excitation Energies (in \si{\wn}) for \ce{FHF-}}
\label{tbl:fhf_FundFreq_results}
\begin{threeparttable}
\begin{tabularx}{.9\textwidth}{l Y Y}
    \toprule
    \toprule
    Method & Bend  & Stretch   \\
    \midrule
    \multicolumn{3}{c}{$R_\text{FF} = 2.271$ \si{\angstrom}}\\
    NEO-MR-SD$_\text{en}$CI\tnote{\emph{a}}       & 1475 & 2071\\
    Reference\tnote{\emph{b}}                     & 1298 & 1658\\
    Multicomponent HCI\tnote{\emph{c}}            & 1836 & 2700\\
    \vspace{1\baselineskip}\\
    \multicolumn{3}{c}{$R_\text{FF} = 2.320$ \si{\angstrom}}\\
    NEO-MR-SD$_\text{en}$CI\tnote{\emph{a}}      & 1511 & 1784\\
    \bottomrule
\end{tabularx}
\begin{tablenotes}\linespread{1}\footnotesize
    \item[\emph{a}]    The electronic basis set for F is aug-cc-pVTZ.
                The electronic and protonic basis sets for H are 0.4742-def2-QZVP* and 8s8p8d8f, respectively.
                The electronic active space is 4 orbitals and 4 electrons.
                The protonic active space is 20 orbitals and 1 proton.
                The orbitals are optimized with NEO-SA-MCSCF averaging over the lowest four vibronic states.
    \item[\emph{b}]    The reference is from an FGH calculation using CCSD and the aug-cc-pVTZ electronic basis set on all atoms.
    \item[\emph{c}]    Result from Ref.~\citenum{NareshBrorsen2021} using a conventional electronic basis set.
\end{tablenotes}
\end{threeparttable}
\end{table}

\subsection{HCN and HNC}
\begin{figure}[htp]
\includegraphics[trim={0, 0, 0, 0},clip,width=3.2in]{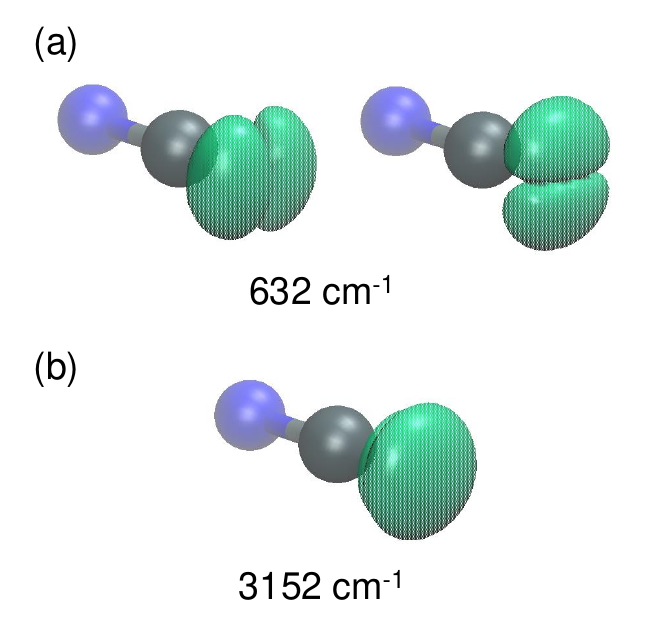}
\caption{Proton vibrational modes for HCN computed using the FGH method in conjunction with conventional electronic CCSD. (a) Degenerate fundamental bend modes and (b) fundamental stretch mode.}
\label{fig:hcn_modes}
\
\end{figure}

The HCN and HNC isomers have similar electronic structures. 
Both systems feature a terminal proton with low-lying degenerate bend modes and a high-frequency stretch mode (Figure \ref{fig:hcn_modes}) that present a challenge for multicomponent methods.\cite{PavosevicTaoCulpittZhaoLiSHS2020,NareshBrorsen2021}
The electronic basis set for carbon and nitrogen is cc-pVTZ.\cite{Dunning1989} A minimal active space was used to describe the protonic motion for both systems, including the H--X (X = C,N) $\sigma$ and $\sigma^*$ orbitals and 2 electrons.
The electronic basis set for the hydrogen is 0.1393-def2-QZVP*.
The basis set scaling factor, $\gamma$, optimized for the HCN ground state protonic density is much lower than the $\gamma$ optimized for \ce{FHF-} and \ce{HeHHe+} by the same procedure.
As the proton is more delocalized in the HCN system, the electronic basis set needed to describe this delocalization must be more diffuse.
Using this electronic basis set for the HNC system without further optimization will demonstrate its transferability to other similar systems with a terminal proton.
The protonic basis set is 8s8p8d8f.

\begin{figure}[htp]
\centering
\includegraphics[width=.99\linewidth]{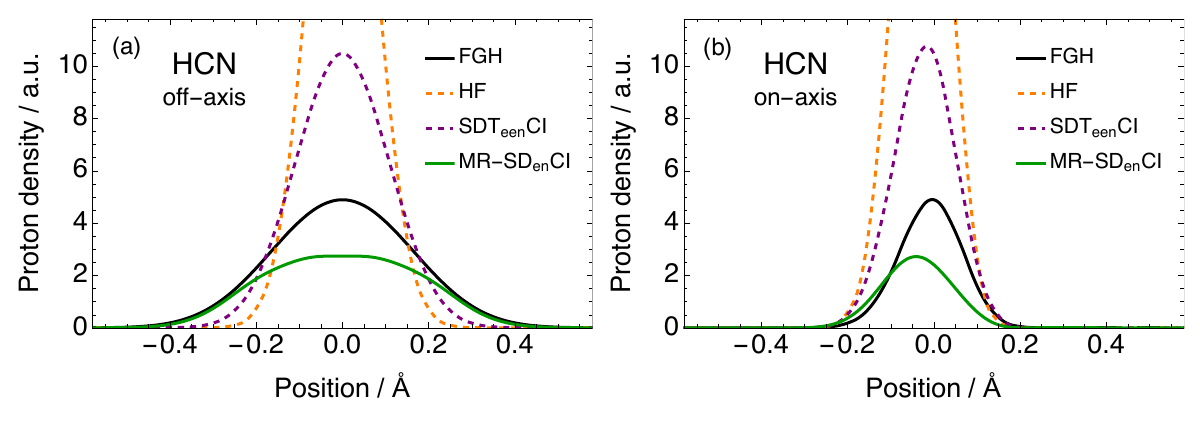}
\caption{Ground state protonic density (a) off-axis and (b) on-axis for HCN computed with NEO-MR-SD$_\text{en}$CI, NEO-SD$_\text{en}$CI, and NEO-HF methods compared to the reference FGH results. 
        The C and N atoms are located at 1.065 \si{\angstrom} and 2.218 \si{\angstrom} along the z-axis, and
        the H atom is placed at the origin.
        The electronic basis set for C and N is cc-pVTZ.
        The electronic and protonic basis sets for H are $0.1393$-def2-QZVP* and 8s8p8d8f, respectively, and are placed at the origin.
        The electronic active is space is 2 orbitals and 2 electrons, and the protonic active space is 16 orbitals and 1 proton.
        The reference FGH calculation uses CCSD and the cc-pVTZ electronic basis set for all atoms.
        The grid spans the range of -0.7935 to 0.8464 \si{\angstrom} around the origin.
        The off-axis plot is at $\text{z}=-0.0969$ \si{\angstrom} for NEO-MR-SD$_\text{en}$CI but at $\text{z} = 0.0$ \si{\angstrom} for all other methods.
        The NEO-HF maximum protonic density is 24 a.u}
\label{fig:hcn_gs_density_compare}
\end{figure}

The fundamental proton bend and stretch mode excitation energies computed with NEO-MR-SD$_\text{en}$CI for HCN are reported in Table \ref{tbl:hcn_and_hnc_FundFreq_results}.
Single electronic excitations from the NEO-MCSCF reference space, treating the carbon 1s and nitrogen 1s orbitals as frozen-core orbitals, results in a NEO-MR-SD$_\text{en}$CI wavefunction of 315 636 NEO configurations.
The lowest two degenerate excited vibronic states are the fundamental proton bend modes. The bend mode excitation energy is only 46 \si{\wn} greater than the FGH reference excitation energy. 
The fourth through seventh vibronic states are overtone vibrations. 
The eighth vibronic state is the fundamental proton stretch mode. 
The stretch mode excitation energy is only 184 \si{\wn} lower than the FGH reference excitation energy.
Figure \ref{fig:hcn_gs_density_compare} shows the ground state protonic density computed with the NEO-MR-SD$_\text{en}$CI, NEO-SDT$_\text{een}$CI, NEO-HF, and reference FGH methods.
The NEO-MR-SD$_\text{en}$CI ground state protonic density is more delocalized than both the FGH and NEO-SDT$_\text{een}$CI proton densities.
It also exhibits a slight shift for the on-axis protonic density, most likely due to the exclusion of electron-electron excitations into the virtual space.
This shifted protonic density has been reported by other methods.\cite{FeldmannMuoloBaiardiReiher2022}
Nevertheless, these NEO protonic densities are much more accurate than those obtained from NEO-HF using the same basis sets, which has a maximum protonic density of 24 a.u.

Also included in Table \ref{tbl:hcn_and_hnc_FundFreq_results} are previous results computed with other multicomponent methods.
The NEO-TDHF method performs the best, but it does so with a much too localized ground state protonic density corresponding to NEO-HF.
The NEO-EOM-CCSD and multicomponent HCI methods overestimate the proton bend mode by over 1000 \si{\wn} and the proton stretch mode by over 500 \si{\wn}.  Thus, the NEO-MR-SD$_\text{en}$CI method produces the best results obtained by a multicomponent wavefunction method to date for this system. As mentioned above, however, these other methods may improve significantly using the revised electronic basis sets developed herein.


\begin{table}
\let\TPToverlap=\TPTrlap
\centering
\caption{Fundamental Vibrational Excitation Energies (in \si{\wn}) for HCN and HNC}
\label{tbl:hcn_and_hnc_FundFreq_results}
\begin{threeparttable}
\begin{tabularx}{.9\textwidth}{l Y Y}
    \toprule
    \toprule
    Method & Bend  & Stretch   \\
    \midrule
    \multicolumn{3}{c}{\textbf{HCN}\tnote{\emph{a}}}\\
    NEO-MR-SD$_\text{en}$CI\tnote{\emph{b,c}}  &  678 &  2968 \\
    Reference\tnote{\emph{d}}              &  632 &  3152 \\
    NEO-TDHF\tnote{\emph{e}}               &  685 &  3281 \\
    NEO-EOM-CCSD\tnote{\emph{f}}           & 1749 &  3768 \\
    Multicomponent HCI\tnote{\emph{g}}     & 1686 &  3615 \\
    \vspace{1\baselineskip}\\
    \multicolumn{3}{c}{\textbf{HNC}\tnote{\emph{h}}}\\
    NEO-MR-SD$_\text{en}$CI\tnote{\emph{b,i}}   &  791 & 3422\\
    NEO-MR-SD$_\text{en}$CI\tnote{\emph{b,j}}   &  550 & 3286\\
    Reference\tnote{\emph{d}}                   &  439 & 3547\\
    NEO-TDHF\tnote{\emph{e}}                    &  631 & 3721\\
    \bottomrule 
\end{tabularx}
\begin{tablenotes}\linespread{1}\footnotesize
    \item[\emph{a}] The C--N distance is 1.153 \si{\angstrom}.
                    The H basis function center is 1.065 \si{\angstrom} from the C atom.
    \item[\emph{b}] The electronic basis set for C and N is cc-pVTZ.
                    The electronic and protonic basis sets for H are 0.1393-def2-QZVP* and 8s8p8d8f, respectively, unless otherwise specified. The orbitals are optimized with NEO-CASSCF.
    \item[\emph{c}] The electronic active is space is 2 orbitals and 2 electrons.
                    The protonic active space is 16 orbitals and 1 proton. 
    \item[\emph{d}]    The reference is from an FGH calculation using CCSD and the cc-pVTZ electronic basis set on all atoms.
    \item[\emph{e}]    Result from Ref.~\citenum{YuPavosevicSHS2020} using a conventional electronic basis set with the cc-pVDZ electronic basis set on C and N.
    \item[\emph{f}]    Result from Ref.~\citenum{PavosevicTaoCulpittZhaoLiSHS2020} using a conventional electronic basis set.
    \item[\emph{g}]    Result from Ref.~\citenum{NareshBrorsen2021} using a conventional electronic basis set.
    \item[\emph{h}]    The C--N distance is 1.168 \si{\angstrom}.
                The H basis function center is 0.994 \si{\angstrom} from the N atom.
    \item[\emph{i}]    The electronic active space is 2 orbitals and 2 electrons. The protonic active space is 87 orbitals and 1 proton.
    \item[\emph{j}]    Result using 8s8p8d8f8g protonic basis set. 
                       The electronic active space is 2 orbitals and 2 electrons.
                       The protonic active space is 116 orbitals and 1 proton.
\end{tablenotes}
\end{threeparttable}
\end{table}

\begin{figure}[htp]
\centering
\includegraphics[width=.99\linewidth]{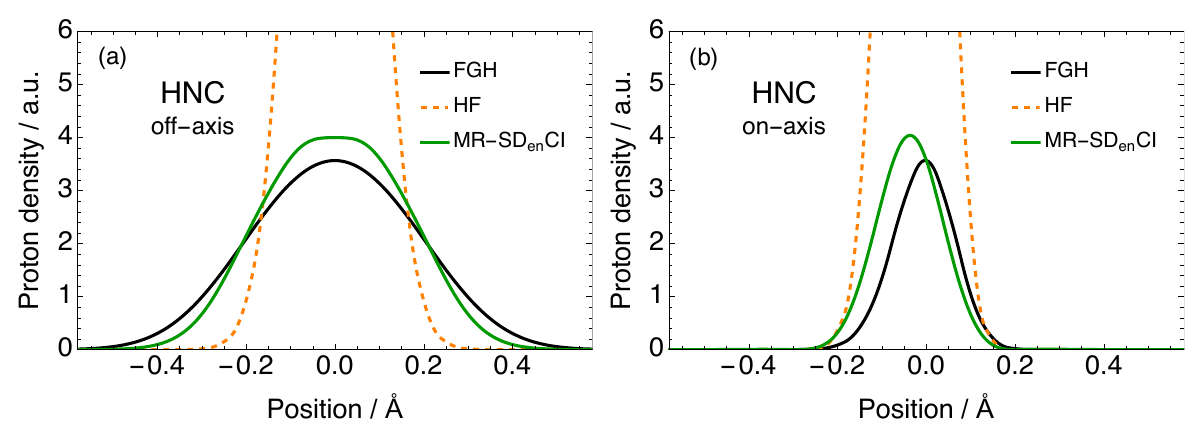}
\caption{Ground state protonic density (a) off-axis and (b) on-axis for HNC computed with NEO-MR-SD$_\text{en}$CI and NEO-HF methods compared to the reference FGH results. 
        The N and C atoms are located at 0.994 \si{\angstrom} and 2.162 \si{\angstrom} along the z-axis, and
        the H atom is placed at the origin.
        The electronic basis set for C and N is cc-pVTZ.
        The electronic and protonic basis sets for H are $0.1393$-def2-QZVP* and 8s8p8d8f, respectively, and are placed at the origin.
        The electronic active is space is 2 orbitals and 2 electrons, and the protonic active space is 87 orbitals and 1 proton.
        The reference FGH calculation uses CCSD and the cc-pVTZ electronic basis set for all atoms.
        The grid spans the range of -0.7935 to 0.8464 \si{\angstrom} around the origin.
        The off-axis plot is at $\text{z}=-0.0485$ \si{\angstrom} for NEO-MR-SD$_\text{en}$CI but at $\text{z} = 0.0$ \si{\angstrom} for all other methods.
        The NEO-HF maximum protonic density is 23 a.u.}
\label{fig:hnc_gs_density_compare}
\end{figure}

In addition, Table \ref{tbl:hcn_and_hnc_FundFreq_results} provides the fundamental proton bend and stretch mode excitation energies computed with NEO-MR-SD$_\text{en}$CI for HNC. 
The same electronic basis set for hydrogen is used for HNC as was used for HCN without further optimization of the $\gamma$ scaling factor.
Using the same protonic basis set as HCN, 8s8p8d8f, the NEO-MR-SD$_\text{en}$CI wavefunction is composed of 193 140 NEO configurations.
The proton stretch mode excitation energy is 125 \si{\wn} below the FGH result, which is satisfactory and consistent with the results for HCN, and the ground state protonic density is in good agreement with the FGH reference{\color{revisions}, particularly in comparison to the NEO-HF ground state protonic density} (Figure \ref{fig:hnc_gs_density_compare}).
However, the proton bend mode excitation energy is overestimated.
Using the larger 8s8p8d8f8g even-tempered basis set, the NEO-MR-SD$_\text{en}$CI wavefunction is composed of 420 848 NEO configurations, and the bend mode excitation energy is 111 \si{\wn} above the reference, an improvement of 241 \si{\wn}, but the stretch mode excitation energy does not agree as well with the FGH reference.
This analysis highlights the challenges faced in describing low-frequency vibrations with multicomponent methods and the need for more robust and extendable electronic and protonic basis sets, as excitation energies are sensitive to the choice of basis set.

{\color{revisions}
Interestingly, we found that the NEO-SD$_\text{en}$CI method also produces accurate proton vibrational excitation energies for the molecular systems studied in this work (Table S4), indicating that these systems most likely do not have strong multireference character.
However, application of the NEO-MR-SD$_\text{en}$CI method to hydrogen tunneling systems\cite{SteinMalbonSHS2025} shows that the NEO-MRCI method can be used to compute accurate tunneling splittings for systems with strong multireference character.\cite{YuSHS2020,DickinsonYuSHS2023}
}


\section{Conclusions} \label{sec:conclusions}


This work presents an implementation of multicomponent MRCI within the NEO framework.
The NEO-MR-SDT$_\text{een}$CI wavefunctions and fundamental proton bend excitation energy agree nearly exactly with the NEO-FCI results and numerically exact grid-based reference for the \ce{HeHHe+} system.
The corresponding overtone and combination vibrational modes are also in excellent agreement with the grid-based reference.
For larger molecular systems, including \ce{FHF-}, HCN, and HNC, the NEO-MR-SD$_\text{en}$CI method provides a reasonably accurate description of the protonic density and fundamental proton vibrational excitation energies.
Thus, the NEO-MRCI approach offers the best description of excited vibronic states in multicomponent wavefunction calculations to date.
The high-level NEO-MRCI approach may also assist in identifying suitable NEO configurations to use in conjunction with the recently developed NEO time-dependent CI (NEO-TDCI) approach for nuclear-electronic quantum dynamics simulations.\cite{GarnerUpadhyayLiSHS2024}

A broader impact of this work arises from {\color{revisions}the clear illustration of the necessity of revising electronic basis sets for use in excited-state multicomponent calculations and the fundamental insights leading to an effective strategy for designing such basis sets.}
Conventional electronic basis sets are not appropriate when the atomic nucleus is no longer a point charge, as such basis sets  are inadequate for describing the electronic structure associated with a delocalized nuclear density {\color{revisions}and do not provide a balanced description of ground and excited vibronic states, favoring the ground state.
Electronic basis sets that provide a balanced description of ground and excited vibronic states are essential for attaining accurate vibronic excitation energies and wavefunctions.}
This finding motivates the development of electronic and accompanying protonic basis sets specifically designed for multicomponent calculations.
For now, the prescription for revising conventional electronic basis sets presented in this work is a viable option that has been shown to be accurate and practical.
These types of revised electronic basis sets are expected to improve all multicomponent methods and will be explored further in future work.

\begin{acknowledgement}

This work was supported by the National Science Foundation
Grant No. CHE-2415034.
This work used Expanse at the San Diego Supercomputer Center through allocation TG-MCB120097 from the Extreme Science and Engineering Discovery Environment (XSEDE),\cite{ExpanseCite} which was supported by National Science Foundation Grant No. 1548562.
This work used Expanse at the San Diego Supercomputer Center through allocation TG-MCB120097 from the Advanced Cyberinfrastructure Coordination Ecosystem: Services \& Support (ACCESS) program, which is supported by U.S. National Science Foundation grants \#2138259, \#2138286, \#2138307, \#2137603, and \#2138296.\cite{AccessCite}
The authors thank Dr.\ Eno Paenurk and Professor Xiaosong Li for insightful discussions.

\end{acknowledgement}

\begin{suppinfo}

Additional equations associated with NEO-CI, NEO-CASSCF, and NEO-RASSCF; additional details about algorithms; pseudocode; additional comparisons of methods; additional protonic density figure; revised electronic basis sets. 

\end{suppinfo}

\bibliography{neo_multirefci_1}

\end{document}


\newpage
\tableofcontents

\newpage

%
%
\section{NEO-CI Density Matrix Expressions}
The NEO multiconfigurational wavefunction is defined as
\begin{equation}
    \Psi^\text{NEO} (\mathbf{R}^c) = \sum_\mu c_\mu (\mathbf{R}^c) \psi_\mu^\text{NEO} (\mathbf{R}^c) ,
\end{equation}
where each NEO configuration $\Psi_\mu^\text{NEO}$ is the product of an electronic and a nuclear determinant:
\begin{equation}
    \psi_\mu^\text{NEO} = \lvert \Phi_{i(\mu)}^\text{e}\rangle \lvert \Phi_{I(\mu)}^\text{n} \rangle.
\end{equation}

The one- and two-particle reduced density matrices for the NEO-CI wavefunction are
\begin{equation}
	D_{pq}^\text{e} = \sum_\mu \sum_\nu c_\mu c_\nu \gamma_{pq}^{\mu\nu}
\end{equation}
\begin{equation}
	P_{pqrs}^\text{e} = \sum_\mu \sum_\nu c_\mu c_\nu \Gamma_{pqrs}^{\mu\nu}
\end{equation}
\begin{equation}
	P_{pqPQ}^\text{en} = \sum_\mu \sum_\nu c_\mu c_\nu \gamma_{pq}^{\mu\nu} \gamma_{PQ}^{\mu\nu}
\end{equation}
\begin{equation}
	D_{PQ}^\text{n} = \sum_\mu \sum_\nu c_\mu c_\nu \gamma_{PQ}^{\mu\nu}
\end{equation}
\begin{equation}
	P_{PQRS}^\text{n} = \sum_\mu \sum_\nu c_\mu c_\nu \Gamma_{PQRS}^{\mu\nu}
\end{equation}
where
\begin{subequations}
    \begin{equation}
        \gamma_{pq}^{\mu\nu} = \langle \psi_\mu^\text{NEO} \vert \hat{E}_{pq} \vert \psi_\nu^\text{NEO} \rangle
    \end{equation}
    \begin{equation}
        \Gamma_{pqrs}^{\mu\nu} = \langle \psi_\mu^\text{NEO} \vert \hat{E}_{pq}\hat{E}_{rs} - \delta_{qr}\hat{E}_{ps} \vert \psi_\nu^\text{NEO} \rangle
    \end{equation}
\end{subequations}
are the electronic one- and two- particle coupling coefficients, and
\begin{subequations}
    \begin{equation}
        \gamma_{PQ}^{\mu\nu} = \langle \psi_\mu^\text{NEO} \vert \hat{E}_{PQ} \vert \psi_\nu^\text{NEO} \rangle
    \end{equation}
    \begin{equation}
        \Gamma_{PQRS}^{\mu\nu} = \langle \psi_\mu^\text{NEO} \vert \hat{E}_{PQ}\hat{E}_{RS} - \delta_{QR}\hat{E}_{PS} \vert \psi_\nu^\text{NEO} \rangle
    \end{equation}
\end{subequations}
are the nuclear one- and two- particle coupling coefficients.
In these expressions $\hat{E}_{pq}$ and $\hat{E}_{PQ}$ are defined as
\begin{subequations}
\begin{equation}
    \hat{E}_{pq} \equiv \sum_\sigma \hat{a}^\dagger_{p\sigma}\hat{a}_{q\sigma}
\end{equation}
\begin{equation}
    \hat{E}_{PQ} \equiv \sum_\sigma \hat{a}^\dagger_{P\sigma}\hat{a}_{Q\sigma}
    \label{eqn:epq_nuc}
\end{equation}
\end{subequations}
where the spin is $\sigma$. In our implementation, the nuclei are treated as high spin, so Eq.~\ref{eqn:epq_nuc} is simply 
\begin{equation}
    \hat{E}_{PQ} = \hat{a}^\dagger_{P}\hat{a}_{Q}
\end{equation}

\clearpage
\newpage

\section{NEO Generalized Fock Matrix Expressions}

In all expressions, we use the following orbital indexing conventions: lower-case letters indicate electronic orbitals and upper-case letters indicate protonic orbitals; 
$i$, $j$, $k$, … indicate inactive occupied orbitals; 
$t$, $u$, $v$,… indicate active occupied orbitals; 
$a$, $b$, $c$, … indicate virtual orbitals; 
and $p$, $q$, $r$, … indicate general orbitals.
The two-electron orbitals are denoted $\left( pq \vert rs \right)$, the two-nucleus orbitals are denoted $\left( PQ \vert RS \right)$, and the mixed-particle, electron-nucleus integrals are denoted $\left( pq \vert PQ \right)$.

The NEO generalized electronic Fock matrix has non-zero elements
\begin{align}
	F_{iq}^{\text{NEO},\text{e}} &= 2(F_{iq}^{\text{NEO},\text{I},\text{e}} + F_{iq}^{\text{NEO},\text{A},\text{e}})\\
	F_{tq}^{\text{NEO},\text{e}} &= \sum_u D_{tu}^\text{e} F_{qu}^{\text{NEO},\text{I},\text{e}} + 2 \sum_{uvx} P_{tuvx}^\text{e} \left(qu \vert vx\right) - \sum_{vUV} P_{tvUV}^\text{en} \left( qv \vert UV \right)
\end{align}
where $\mathbf{D}^\text{e}$ and $\mathbf{P}^\text{e}$ are the electronic one- and two-particle reduced density matrices, respectively; $\mathbf{P}^\text{en}$ is the mixed electronic-nuclear two-particle reduced density matrix; and $\mathbf{F}^{\text{NEO},\text{I},\text{e}}$ and $\mathbf{F}^{\text{NEO},\text{A},\text{e}}$ are the inactive and active electronic Fock matrices, respectively.
The inactive and active electronic Fock matrices have elements
\begin{align}
	F_{pq}^{\text{NEO},\text{I},\text{e}}  &= h^\text{e}_{pq} + \sum_i \left[ 2\left(pq \vert ii \right) - \left(pi \vert qi\right) \right] - \sum_I \left( pq \vert II \right)\\
	F_{pq}^{\text{NEO},\text{A},\text{e}} &= \sum_{tu} D_{tu}^\text{e} \left[ \left( pq \vert tu \right) - \frac{1}{2} \left( pt \vert qt \right) \right] - \sum_{TU} D_{TU}^\text{n} \left( pq \vert TU \right)
\end{align}

The NEO generalized nuclear Fock matrix has non-zero elements
\begin{align}
	F_{IQ}^{\text{NEO},\text{n}}  & = F_{IQ}^{\text{NEO},\text{I},\text{n}} + F_{IQ}^{\text{NEO},\text{A},\text{n}}\\
	F_{TQ}^{\text{NEO},\text{n}} & = \sum_{U} D_{TU}^\text{n} F_{QU}^{\text{NEO},\text{I},\text{n}} + 2\sum_{UVX} P_{TUVX}^\text{n} \left(QU \vert VX \right) - \sum_{Vuv} P_{TVuv}^\text{en} \left( QV \vert uv \right)
\end{align}
where $\mathbf{D}^\text{n}$ and $\mathbf{P}^\text{n}$ are the nuclear one- and two-particle reduced density matrices, respectively; $\mathbf{P}^\text{en}$ is the mixed electronic-nuclear two-particle reduced density matrix; and $\mathbf{F}^{\text{NEO},\text{I},\text{n}}$ and $\mathbf{F}^{\text{NEO},\text{A},\text{n}}$ are the inactive and active nuclear Fock matrices, respectively.
The inactive and active nuclear Fock matrices have elements
\begin{align}
	F_{PQ}^{I,\text{n}}  & = h^\text{n}_{PQ} + \sum_I \left[ \left( PQ \vert II \right) - \left( PI \vert QI \right) \right] - \sum_i \left( PQ \vert ii \right)\\
	F_{PQ}^{A,\text{n}} & = \sum_{TU} D_{TU}^\text{n} \left[ \left( PQ \vert TU \right) - \left( PT \vert QT \right) \right] - \sum_{tu} D_{tu}^\text{e} \left( PQ \vert tu \right)
\end{align}

%
%
\newpage
\section{NEO-CASSCF Orbital Hessian Expressions} \label{sec:neo_orb_hess_ee}
Below are the matrix element expressions of the NEO-MCSCF electronic orbital Hessian.
The NEO-MCSCF nuclear orbital Hessian, dependent upon $X^\text{n}_{PQ}$ is completely analogous.

\begin{align}
\begin{split}
\frac{1}{2}\frac{\partial^2E}{\partial X^\text{e}_{it} \partial X^\text{e}_{ju}} = & 
		2 \sum_{vx} \left[P_{utvx}^\text{e}\left(vx\vert ij\right) + \left(P_{uxvt}^\text{e} + P_{uxtv}^\text{e}\right)\left(vi\vert xj\right)\right]\\
		&+ \sum_{v}\left(\delta_{tv} - D_{tv}^\text{e}\right)\left[4\left(vi\vert uj\right) - \left(ui\vert vj\right) - \left(uv\vert ij\right)\right]\\
		&+ \sum_{v}\left(\delta_{uv} - D_{uv}^\text{e}\right)\left[4\left(vj\vert ti\right) - \left(tj\vert vi\right) - \left(tv\vert ij\right)\right]\\
		&+ D_{tu}^\text{e} F_{ij}^{\text{NEO},\text{I},\text{e}} + \delta_{ij}\left(2F_{tu}^{\text{NEO},\text{I},\text{e}} + 2F_{tu}^{\text{NEO},\text{A},\text{e}} - F_{tu}^{\text{NEO},\text{e}}\right)\\
		& + 2\delta_{tu}\left(F_{ij}^{\text{NEO},\text{I},\text{e}} + F_{ij}^{\text{NEO},\text{A},\text{e}}\right) - \sum_{TU}\left(ji\vert TU\right)P_{tuTU}^\text{en}
\end{split}\\
    \begin{split}
	\frac{1}{2}\frac{\partial^2E}{\partial X^\text{e}_{it} \partial X^\text{e}_{ja}} = &
			\sum_v\left(2\delta_{tv} - D_{tv}^\text{e} \right)\left[4\left(aj \vert vi\right) - \left(av \vert ij\right) - \left(ai \vert vj\right)\right] \\
			& + 2\delta_{ij}\left(F_{at}^{\text{NEO},\text{I},\text{e}} + F_{at}^{\text{NEO},\text{A},\text{e}}\right) - \frac{1}{2}\delta_{ij}F_{ta}^{\text{NEO},\text{e}}
    \end{split}\\
    \begin{split}
    \frac{1}{2}\frac{\partial^2E}{\partial X^\text{e}_{it} \partial X^\text{e}_{ua}} = &
			-2 \sum_{vx} \left[P_{tuvx}^\text{e}\left(ai \vert vx\right) + \left(P_{tvux}^\text{e} + P_{tvxu}^\text{e}\right)\left(ax \vert vi\right)\right] \\
			&+ \sum_v D_{uv}^\text{e}\left[4\left(av \vert ti\right) - \left(ai \vert tv\right) - \left(at \vert vi\right)\right] \\
			&- D_{tu}^\text{e} F_{ai}^{\text{NEO},\text{I},\text{e}} + \delta_{tu}\left(F_{ai}^{\text{NEO},\text{I},\text{e}} + F_{ai}^{\text{NEO},\text{A},\text{e}}\right) \\
			&+ \sum_{TU}\left(ai \vert TU\right)P_{tuTU}^\text{en}
    \end{split}\\
\begin{split}
    \frac{1}{2}\frac{\partial^2E}{\partial X^\text{e}_{ia} \partial X^\text{e}_{jb}} = &
			2\left[4\left(ai \vert bj\right) - \left(ab \vert ij\right) - \left(aj \vert bi\right)\right] + 2\delta_{ij}\left(F_{ab}^{\text{NEO},\text{I},\text{e}} + F_{ab}^{\text{NEO},\text{A},\text{e}}\right)\\
			&- 2\delta_{ab}\left(F_{ij}^{\text{NEO},\text{I},\text{e}} + F_{ij}^{\text{NEO},\text{A},\text{e}}\right) 
\end{split}\\
\begin{split}
    \frac{1}{2}\frac{\partial^2E}{\partial X^\text{e}_{ia} \partial X^\text{e}_{tb}} = &
			\sum_v D_{tv}^\text{e}\left[4\left(ai \vert bv\right) - \left(av \vert bi\right) - \left(ab \vert vi\right)\right] \\
			&- \delta_{ab}\left(F_{ti}^{\text{NEO},\text{I},\text{e}} + F_{ti}^{\text{NEO},\text{A},\text{e}}\right) - \frac{1}{2}\delta_{ab}F_{ti}^{\text{NEO},\text{e}}
\end{split}
\end{align}
\begin{align}
\begin{split}
    \frac{1}{2}\frac{\partial^2E}{\partial X^\text{e}_{ta} \partial X^\text{e}_{ub}} = &
			\sum_{vx}\left[P_{tuvx}^\text{e}\left(ab \vert vx\right) + \left(P_{txvu}^\text{e} + P_{txuv}^\text{e}\right)\left(ax \vert bv\right)\right] \\
			&+ D_{tu}^\text{e} F_{ab}^{\text{NEO},\text{I},\text{e}} - \delta_{ab}F_{tu}^{\text{NEO},\text{e}} - \sum_{TU}\left(ba \vert TU\right)P_{tuTU}^\text{en}
\end{split}
\end{align}

\clearpage
\newpage
\section{NEO-RASSCF Orbital Hessian Expressions}
The restricted active space SCF (RASSCF) method requires the calculation of additional Hessian matrix elements for the contributions of rotations in and out of the different RAS spaces.
The expressions for the two types of terms (diagonal and off-diagonal) are presented below.

\begin{align}
\begin{split}
    \frac{1}{2}\frac{\partial^2E}{\partial X^\text{e}_{tu} \partial X^\text{e}_{tu}} = 
    & D^\text{e}_{tt}F^{\text{NEO},\text{I},\text{e}}_{uu} - 2D^\text{e}_{ut}F^{\text{NEO},\text{I},\text{e}}_{tu} + D^\text{e}_{uu}F^{\text{NEO},\text{I},\text{e}}_{tt} - F^{\text{NEO},\text{e}}_{uu} - F^{\text{NEO},\text{e}}_{tt}\\
    &+ 2\sum_{vx}\left[P^\text{e}_{ttvx}\left(uu\vert vx\right) + \left(P^\text{e}_{txvt} + P^\text{e}_{tvtx}\right)\left(uv\vert ux\right)\right] \\
    &- 2\sum_{vx}\left[P^\text{e}_{tuvx}\left(ut\vert vx\right) + \left(P^\text{e}_{txvu} + P^\text{e}_{tvux}\right)\left(uv\vert tx\right)\right] \\
    &- 2\sum_{vx}\left[P^\text{e}_{utvx}\left(tu\vert vx\right) + \left(P^\text{e}_{uxvt} + P^\text{e}_{uvtx}\right)\left(tv\vert ux\right)\right] \\
    &+ 2\sum_{vx}\left[P^\text{e}_{uuvx}\left(tt\vert vx\right) + \left(P^\text{e}_{uxvu} + P^\text{e}_{uvux}\right)\left(tv\vert tx\right)\right] \\
    &+ \sum_{TU}\left[ 2P^\text{en}_{utTU}\left(ut\vert TU\right) - P^\text{en}_{ttTU}\left(uu\vert TU\right) - P^\text{en}_{uuTU}\left(tt\vert TU\right)\right]
\end{split}\\
\begin{split}
    \frac{1}{2}\frac{\partial^2E}{\partial X^\text{e}_{tu} \partial X^\text{e}_{vx}} = 
    & D^\text{e}_{tv}F^{\text{NEO},\text{I},\text{e}}_{ux} - D^\text{e}_{uv}F^{\text{NEO},\text{I},\text{e}}_{tx} - D^\text{e}_{tx}F^{\text{NEO},\text{I},\text{e}}_{uv} + D^\text{e}_{ux}F^{\text{NEO},\text{I},\text{e}}_{tv}\\
    &+ 2\sum_{wz}\left[P^\text{e}_{vtwz}\left(ux\vert wz\right) + \left(P^\text{e}_{vzwt} + P^\text{e}_{vztw}\right)\left(uw\vert xz\right)\right] \\
    &- 2\sum_{wz}\left[P^\text{e}_{vuwz}\left(tx\vert wz\right) + \left(P^\text{e}_{vzwu} + P^\text{e}_{vzuw}\right)\left(tw\vert xz\right)\right] \\
    &- 2\sum_{wz}\left[P^\text{e}_{xtwz}\left(uv\vert wz\right) + \left(P^\text{e}_{xzwt} + P^\text{e}_{xztw}\right)\left(uw\vert vz\right)\right] \\
    &+ 2\sum_{wz}\left[P^\text{e}_{xuwz}\left(tv\vert wz\right) + \left(P^\text{e}_{xzwu} + P^\text{e}_{xzuw}\right)\left(tw\vert vz\right)\right] \\
    &+ \frac{1}{2}\delta_{uv}\left(F^{\text{NEO},\text{e}}_{tx} + F^{\text{NEO},\text{e}}_{xt}\right) - \frac{1}{2}\delta_{tv}\left(F^{\text{NEO},\text{e}}_{ux} + F^{\text{NEO},\text{e}}_{xu}\right)\\
    &- \frac{1}{2}\delta_{ux}\left(F^{\text{NEO},\text{e}}_{tv} + F^{\text{NEO},\text{e}}_{vt}\right) + \frac{1}{2}\delta_{tx}\left(F^{\text{NEO},\text{e}}_{uv} + F^{\text{NEO},\text{e}}_{vu}\right) \\
    &+ \sum_{TU}\left(xt\vert TU\right)P^\text{en}_{uvTU} - \sum_{TU}\left(xu\vert TU\right)P^\text{en}_{tvTU}\\
    &- \sum_{TU}\left(vt\vert TU\right)P^\text{en}_{uxTU} + \sum_{TU}\left(vu\vert TU\right)P^\text{en}_{txTU}
\end{split}
\end{align}

\clearpage
\newpage
\section {NEO Mixed-particle Hessian Expressions}
Below are mixed electronic-nuclear orbital Hessian matrix element expressions for multicomponent MCSCF expansions used in this work.

\begin{align}
\begin{split}
    -\frac{1}{2}\frac{\partial^2E}{\partial X^\text{e}_{it} \partial X^\text{n}_{TA}} = &
            4 \sum_U \left(ti\vert AU\right)D^\text{n}_{TU} - 2 \sum_u\sum_U \left(iu\vert AU\right)P^\text{en}_{tuTU}
\end{split}\\
\begin{split}
    -\frac{1}{2}\frac{\partial^2E}{\partial X^\text{e}_{ia} \partial X^\text{n}_{TA}} = &
            4 \sum_U \left(ai\vert AU\right)D^\text{n}_{TU}
\end{split}\\
\begin{split}
    -\frac{1}{2}\frac{\partial^2E}{\partial X^\text{e}_{ta} \partial X^\text{n}_{TA}} = &
            2 \sum_q \sum_Q \left(aq\vert AQ\right) P^\text{en}_{tqTQ}
\end{split}
\end{align}

\clearpage
\newpage
\section{Augmented Hessian}
The augmented Hessian is defined as
\begin{equation}
	\tilde{\textbf{A}} (\lambda) \equiv \begin{pmatrix} 0 &\lambda \textbf{W}^\dagger \\ \lambda \textbf{W} & \textbf{A} \end{pmatrix}
\end{equation}
where $\textbf{W}$ and $\textbf{A}$ are the NEO-MCSCF orbital gradient and orbital Hessian, respectively, and $\lambda$ is a scaling parameter.
The eigenvalue problem
\begin{equation}
	\begin{pmatrix} 0 &\lambda \textbf{W}^\dagger \\ \lambda \textbf{W} & \textbf{A} \end{pmatrix} \begin{pmatrix} 1 \\ \textbf{s}(\lambda) \end{pmatrix} = \zeta \begin{pmatrix} 1 \\ \textbf{s}(\lambda) \end{pmatrix}
\end{equation}
results in two equations:
\begin{subequations}
\begin{align}
	\lambda \textbf{W}^\dagger \textbf{s}(\lambda) & = \zeta \label{eqn:ahEq1}\\
	\lambda \textbf{W} + \textbf{A} \textbf{s}(\lambda) &= \zeta\textbf{s}(\lambda) \label{eqn:ahEq2}
\end{align}
\end{subequations}
Equation \ref{eqn:ahEq1} represents a level-shifted Newton-Raphson solution.

The step vector $\mathbf{s}(\lambda)$ and level-shift $\zeta$ are obtained from diagonalizing the augmented Hessian matrix. Only the lowest root is required, so the Davidson algorithm is used.

\clearpage
\newpage
\section{NEO-CI}
The CI wavefunction is defined by minimum and maximum particle-orbital occupation rules that are established within each orbital space.
Orbital spaces for MCSCF wavefunctions are the familiar doubly-occupied, active-occupied, and virtual-unoccupied spaces. RASSCF wavefuntions have additional spaces: restricted-active-occupied and auxillary-occupied spaces.
MRCI wavefunctions are defined by the maximum number of particles in the reference virtual orbital space.

In the string-based algorithm of Ref. \citenum{IvanicRuedenberg2001}, single and double replacements are generated for each string during the evaluation of $\mathbf{H}\mathbf{v} = \boldsymbol{\sigma}$.
Because not all excitations form valid determinants, this must be checked for each matrix element.
Considerable computation effort is saved if alpha and beta strings are grouped by particle-orbital occupations, which are defined by the expansion, and the list of valid groupings is saved.
If an alpha string group, $G^\alpha_i$, and a beta string group, $G^\beta_j$, satisfy the wavefunction rules, then all alpha and beta strings of these groups form valid determinants.
Practically, this means we need to only check that the groupings, rather than each individual string, satisfy our expansion rules.
Pseudocode outlining the algorithm is presented in Section \ref{sec:si_pseudocode_for_hv}.
NEO configurations are formed from valid electronic and nuclear determinants.
For every CI expansion in this work, there are no restrictions on electronic/nuclear determinant pairings, i.e., all electronic determinants for valid NEO determinants are paired  with all nuclear determinants.

\clearpage
\newpage
\section{Pseudocode for evaluation of $\textbf{H} \textbf{v} = \boldsymbol{\sigma}$} \label{sec:si_pseudocode_for_hv}

In the evaluation of $\textbf{H} \textbf{v} = \boldsymbol{\sigma}$, $\boldsymbol{\sigma}$ is the sum of ten parts: diagonal elements, $\boldsymbol{\sigma}_1$; single alpha replacements, $\boldsymbol{\sigma}_2$; double alpha replacements, $\boldsymbol{\sigma}_3$; single beta replacements, $\boldsymbol{\sigma}_4$;
double beta replacements, $\boldsymbol{\sigma}_5$; single alpha and single beta replacements, $\boldsymbol{\sigma}_6$; single protonic replacements, $\boldsymbol{\sigma}_7$; double protonic replacements, $\boldsymbol{\sigma}_8$;
single alpha and single protonic replacements, $\boldsymbol{\sigma}_9$; and single beta and single protonic replacements, $\boldsymbol{\sigma}_{10}$. Here alpha and beta refer to the spin of the electron.
\[
	\boldsymbol{\sigma} = 	\boldsymbol{\sigma}_1 + \boldsymbol{\sigma}_2 + \boldsymbol{\sigma}_3 + \boldsymbol{\sigma}_4 +
						\boldsymbol{\sigma}_5 + \boldsymbol{\sigma}_6 + \boldsymbol{\sigma}_7 + \boldsymbol{\sigma}_8 +
						\boldsymbol{\sigma}_9 + \boldsymbol{\sigma}_{10}
\]

Figures \ref{fig:si_sigma_1} to \ref{fig:si_sigma_9} provide the pseudocode for the computation of each contribution to $\boldsymbol{\sigma}$.
Alpha and beta electronic orbital replacement contributions are completely analogous to one another; therefore, beta replacement contributions are omitted.
In each pseudocode example, $\bm{\alpha}$ represents an alpha electronic orbital occupation string, $\bm{\beta}$ represents a beta electronic orbital occupation string, $\bm{n}$ represents a nuclear orbital occupation string, and primes are used to indicate replacements in the occupation strings.
Primes signify a replacement of one (') or two (") orbitals in an occupation string, e.g. $\bm{\alpha}' \gets \bm{\alpha}$ is the replacement of one occupied orbital of alpha string $\bm{\alpha}$ with an unoccupied orbital. 

\begin{figure}
\begin{algorithmic}[ht]
\FOR{$\bm{\alpha}' \gets \bm{\alpha}$}
	\STATE \text{Get electron-orbital occupation space of $\bm{\alpha}'$, $G^\alpha_j$}
	\STATE \text{Get $G^\beta_k$ that forms a valid determinant with $G^\alpha_i$ and $G^\alpha_j$}
	\FOR{$\bm{\beta} \in G^\beta_k$}
		\FOR{$\bm{n}$}
			\STATE \text{$\sigma(\bm{\alpha}, \bm{\beta}, \bm{n}) \mathrel{+}= \langle \bm{\alpha} , \bm{\beta}, \bm{n} \vert \hat{H} \vert \bm{\alpha}', \bm{\beta}, \bm{n} \rangle \cdot v(\bm{\alpha}',\bm{\beta}, \bm{n})$}
		\ENDFOR
	\ENDFOR
\ENDFOR
\end{algorithmic}
\caption{Computation of single replacements in alpha strings.}
\label{fig:si_sigma_1}
\end{figure} 

\begin{figure}
\begin{algorithmic}[ht]
\FOR{$\bm{\alpha}' \gets \bm{\alpha}$}
	\FOR{$\bm{\alpha}'' \gets \bm{\alpha}'$}
		\STATE \text{Get electron-orbital occupation space of $\bm{\alpha}''$, $G^\alpha_j$}
		\STATE \text{Get $G^\beta_k$ that forms a valid determinant with $G^\alpha_i$ and $G^\alpha_j$}
		\FOR{$\bm{\beta} \in G^\beta_k$}
			\FOR{$\bm{n}$}
				\STATE \text{$\sigma(\bm{\alpha}, \bm{\beta}, \bm{n}) \mathrel{+}= \langle \bm{\alpha} , \bm{\beta}, \bm{n} \vert \hat{H} \vert \bm{\alpha}'', \bm{\beta}, \bm{n} \rangle \cdot v(\bm{\alpha}'',\bm{\beta}, \bm{n})$}
			\ENDFOR
		\ENDFOR
	\ENDFOR
\ENDFOR
\end{algorithmic}
\caption{Computation of double replacements in alpha strings.}
\end{figure}

\begin{figure}
\begin{algorithmic}[ht]
\FOR{$\bm{\alpha}' \gets \bm{\alpha}$}
	\STATE \text{Get electron-orbital occupation space of $\bm{\alpha}'$, $G^\alpha_j$}
	\STATE \text{Get $G^\beta_k$ that forms a valid determinant with $G^\alpha_i$}
	\FOR{$\bm{\beta} \in G^\beta_k$}
		\FOR{$\bm{\beta}' \gets \bm{\beta}, \text {where } \bm{\beta}' \in G^\beta_l \text{ and } G^\beta_l \text{ pairs with } G^\alpha_j$}
			\FOR{$\bm{n}$}
				\STATE \text{$ \sigma(\bm{\alpha},\bm{\beta}, \bm{n}) \mathrel{+}= \langle \bm{\alpha}, \bm{\beta}, \bm{n} \vert \hat{H} \vert \bm{\alpha}', \bm{\beta}', \bm{n} \rangle \cdot v(\bm{\alpha}',\bm{\beta}', \bm{n})$}
			\ENDFOR
		\ENDFOR
	\ENDFOR
\ENDFOR
\end{algorithmic}
\caption{Computation of double replacements in alpha (single) and beta (single).}
\end{figure}

\begin{figure}
\begin{algorithmic}[ht]
\FOR{$\bm{n}' \gets \bm{n}$}
	\FOR{all $\{G^\alpha_i, G^\beta_k\}$ pairs}
		\FOR{$\bm{\alpha} \in G^\alpha_i$}
			\FOR{$\bm{\beta} \in G^\beta_k$}
				\STATE \text{$\sigma(\bm{\alpha},\bm{\beta}, \bm{n}) \mathrel{+}= \langle \bm{\alpha}, \bm{\beta}, \bm{n} \vert \hat{H} \vert \bm{\alpha}, \bm{\beta}, \bm{n}' \rangle \cdot v(\bm{\alpha},\bm{\beta},\bm{n}') $}
			\ENDFOR
		\ENDFOR
	\ENDFOR
\ENDFOR
\end{algorithmic}
\caption{Computation of single replacements in nuclear strings.}
\end{figure}

\begin{figure}
\begin{algorithmic}[ht]
\FOR{$\bm{\alpha}' \gets \bm{\alpha}$}
	\STATE \text{Get electron-orbital occupation space of $\bm{\alpha}'$, $G^\alpha_j$}
	\STATE \text{Get $G^\beta_k$ that forms a valid determinant with $G^\alpha_i$ and $G^\alpha_j$}
	\FOR{$\bm{\beta} \in G^\beta_k$}
		\FOR{$\bm{n}' \gets \bm{n}$}
			\STATE \text{$\sigma(\bm{\alpha}, \bm{\beta}, \bm{n}) \mathrel{+}= \langle \bm{\alpha} , \bm{\beta}, \bm{n} \vert \hat{H} \vert \bm{\alpha}', \bm{\beta}, \bm{n}' \rangle \cdot v(\bm{\alpha}',\bm{\beta}, \bm{n}')$}
		\ENDFOR
	\ENDFOR
\ENDFOR
\end{algorithmic}
\caption{Computation of double replacements in alpha (single) and nuclear (single) strings.}
\label{fig:si_sigma_9}
\end{figure}

\clearpage
\newpage

\section{Comparison to other multicomponent methods} \label{sec:si_comparison_to_reiher}
Table \ref{tbl:HeHHe_results_compared_to_other_methods_and_bases} reports the excitation energy for the first excited vibronic state computed with different multicomponent wavefunction methods using a conventional electronic basis set and an electronic basis set that has been modified and scaled. The modified electronic basis set produces significantly lower excitation energies that are in much better agreement with the FGH reference value.

\begin{table}[h]
\let\TPToverlap=\TPTrlap
\centering
\caption{Excitation Energy for First Excited Vibronic State of \ce{HeHHe+} (in \si{\wn}) Computed with Conventional and Modified def2-QZVP Basis Sets for $R_{\text{He---He}} = 1.8$ \si{\angstrom}.}
\label{tbl:HeHHe_results_compared_to_other_methods_and_bases}
\begin{threeparttable}
\begin{tabularx}{.9\textwidth}{l Z}
    \toprule
    \toprule
    Method & $E_1 - E_0$\\
    \midrule
    \multicolumn{2}{c}{\underline{def2-QZVP / PB5G}}\\
    NEO-SD$_\text{en}$CI                 & 1101\\
    NEO-SDT$_\text{een}$CI               & 1223\\
    NEO-MR-SD$_\text{en}$CI\tnote{\emph{a}}     & 1195\\
    NEO-MR-SDT$_\text{een}$CI\tnote{\emph{a}}   & 1184\\
    NEO-FCI                     & 1183\\
    NEAP-DMRG\tnote{\emph{b}}          & 1145\\
    \multicolumn{2}{c}{\underline{0.4404-def2-QZVP*\tnote{\emph{c}} / 8s8p8d8f}}\\
    NEO-SD$_\text{en}$CI                 & 699\\
    NEO-SDT$_\text{een}$CI               & 829\\
    NEO-MR-SD$_\text{en}$CI\tnote{\emph{a}}     & 805\\
    NEO-MR-SDT$_\text{een}$CI\tnote{\emph{a}}   & 786\\
    NEO-FCI                     & 783\\
    \vspace{1\baselineskip}\\
    Reference\tnote{\emph{d}}          & 771\\
    \bottomrule 
\end{tabularx}
\begin{tablenotes}\linespread{1}\footnotesize
    \item[\emph{a}]    The electronic active space is 6 orbitals and 4 electrons.
                The protonic active space is 87 orbitals and 1 proton.
    \item[\emph{b}]    Value obtained from Ref. \citenum{FeldmannMuoloBaiardiReiher2022}. 
                Note that the FGH result from Ref. \citenum{SkonePakSHS2005} used to benchmark the proton bend modes in Ref. \citenum{FeldmannMuoloBaiardiReiher2022} is  actually the proton stretch mode.
    \item[\emph{c}]    Scaling factor for basis set exponents determined via NEO-SDT$_\text{een}$CI, minimizing the RMSE of the proton density compared to the FGH reference.
    \item[\emph{d}]    The reference is from an FGH calculation using FCI and the 6-31G electronic basis set on all atoms.
\end{tablenotes}
\end{threeparttable}
\end{table}

\clearpage
\newpage

\section{The method of optimizing $\gamma$}

Let $\Gamma$ be the ordered set of $k$ values of $\gamma$
\begin{equation}
    \Gamma = \left( \gamma_1, \gamma_2, ..., \gamma_k \right) \label{eqn:set_of_gammas}
\end{equation}
and $P(\gamma)$ to be
\begin{equation}
    P(\gamma) = \sqrt{\frac{\sum\limits_{i=1}^{N}\left( \rho^i_\text{NEO}(\gamma) - \rho^i_\text{FGH} \right)^2}{N}} \label{eqn:pfucn}
\end{equation}
where $N = 32 768$ is the number of grid points, $\rho^i_\text{NEO}$ is the NEO-SDT$_\text{een}$CI ground state proton density at grid point $i$, and $\rho^i_\text{FGH}$ is the FGH ground state proton density at grid point $i$.
$P$ is the root-mean square error (RMSE) between the NEO-SDT$_\text{een}$CI and the FGH ground state proton densities.

Define $S_P$ as
\begin{equation}
    S_P = \{P(\gamma_k) \mid \gamma_k \in \Gamma\} \label{eqn:set_sp}
\end{equation}
Take the minimum of $S_P$ to be $P(\gamma_m)$.
A quadratic function $F(\gamma)$, defined as 
\begin{equation}
    F(\gamma) = a\cdot\gamma^2 + b\cdot\gamma + c \label{eqn:quadfit}
\end{equation}
is fit to three points: $P(\gamma_{m-1})$, $P(\gamma_m)$, $P(\gamma_{m+1})$, where 
$a$, $b$, and $c$ are fitting parameters.
Take $\gamma_\text{new}$ to be the $\gamma$ that minimizes Eq.~\ref{eqn:quadfit}.

Pseudocode outlining the algorithm for optimizing $\gamma$ is presented in Fig.~\ref{fig:gamma_algo}.
\begin{figure}
\centering
\begin{algorithmic}[h]
\STATE $k = 9$
\STATE $\Gamma = (0.1, 0.2, ..., 0.9)$
\WHILE{$true$}
	\STATE $\text{Get } F(\gamma)$
    \STATE $\text{Add } \gamma_\text{new} \text{ to } \Gamma$
    \STATE k = k + 1
    \IF{$P(\gamma_m) = P(\gamma_\text{new})$}
        \STATE $\text{exit}$
    \ENDIF
\ENDWHILE
\end{algorithmic}
\caption{Algorithm to optimize $\gamma$ parameter for electronic basis sets.}
\label{fig:gamma_algo}
\end{figure}

\clearpage
\newpage

\section{Modified and scaled electronic basis sets} \label{sec:ebasis_sets}
Below are the coefficients of the three H electronic basis sets used in this work.
\begin{table}[htp]
\let\TPToverlap=\TPTrlap
\centering
\begin{threeparttable}
\caption{   Exponent Parameters for $\gamma$-def2-QZVP* Electronic Basis Functions.}
\label{tbl:def2qzvp_opt_coeffs_4404}
\begin{tabularx}{.9\textwidth}{Y | Y Y Y}
    \toprule
    \toprule
      & \multicolumn{3}{c}{Exponents}\\
    Basis function     & $\gamma = 0.4742$ & $\gamma = 0.4404$ & $\gamma = 0.1393$\\
    \midrule
    $s$   &   0.873119 &   0.810885 &   0.256485\\
    $s$   &   0.283826 &   0.263596 &   0.083376\\
    $s$   &   0.101468 &   0.094235 &   0.029807\\
    $s$   &   0.038086 &   0.035371 &   0.011188\\
    \midrule
    $p$   &   1.086866 &   1.009397 &   0.319276\\
    $p$   &   0.397380 &   0.369055 &   0.116733\\
    $p$   &   0.138466 &   0.128597 &   0.040676\\
    \midrule
    $d$   &   0.977800 &   0.908105 &   0.287237\\
    $d$   &   0.313920 &   0.291545 &   0.092217\\
    \midrule
    $f$   &   0.662457 &   0.615239 &   0.194602\\
    \bottomrule
\end{tabularx}
\end{threeparttable}
\end{table}

\newpage

\section{Protonic density of proton stretch mode of \ce{FHF-}}
\begin{figure}[htp]
\centering
\includegraphics[width=3.25in]{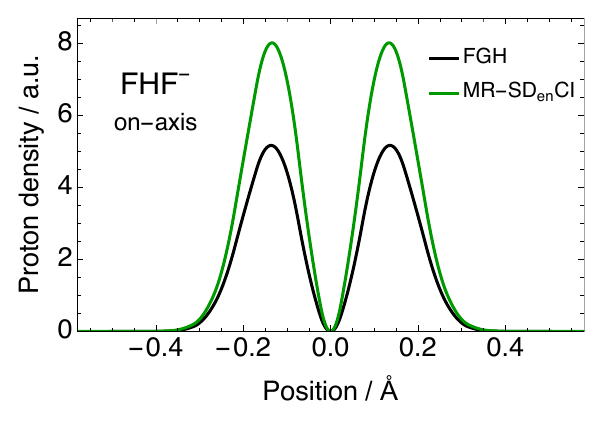}
\caption{Excited state protonic density corresponding to the proton stretch mode of \ce{FHF-} computed with NEO-MR-SD$_\text{en}$CI compared to the reference FGH results. 
The distance between the fixed F atoms is 2.2714 \si{\angstrom} with the midpoint at the origin. 
The electronic basis set for F is aug-cc-pVTZ.
The electronic and protonic basis sets for H are $0.4742$-def2-QZVP* and 8s8p8d8f, respectively, and are placed at the origin.
The electronic active space is 4 orbitals and 4 electrons, and the protonic active space is 20 orbitals and 1 proton.
The reference FGH calculation uses CCSD and the aug-cc-pVTZ electronic basis set for all atoms.
The grid spans the range of -0.7275 to 0.7760 \si{\angstrom} around the origin.}
\label{fig:fhf_stre_density_compare}
\end{figure}

\newpage

\section{Ground state protonic density for HNC with 8s8p8d8f and 8s8p8d8f8g protonic basis sets}
\begin{figure}[htp]
\centering
\includegraphics[width=.99\linewidth]{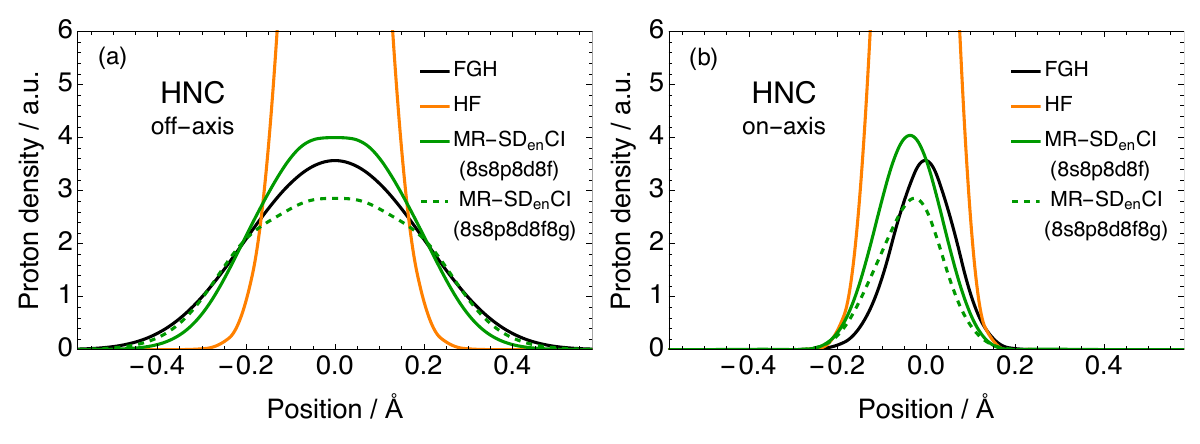}
\caption{Ground state protonic density (a) off-axis and (b) on-axis for HNC computed with NEO-MR-SD$_\text{en}$CI and NEO-HF methods compared to the reference FGH results.
        The N and C atoms are located at 0.994 \si{\angstrom} and 2.162 \si{\angstrom} along the z-axis, and the H atom is placed at the origin.
        The electronic basis set for C and N is cc-pVTZ.
        The electronic basis set for H is 0.1393-def2-QZVP*.
        The protonic basis set is 8s8p8d8f (solid) or 8s8p8d8f8g (dashed).
        The electronic active space is 2 orbitals and 2 electrons.
        The protonic active space is 87 orbitals (solid) or 116 orbitals (dashed) and 1 proton.
        The reference FGH calculation uses CCSD and the cc-pVTZ electronic basis set for all atoms.
        The grid spans the range of -0.7935 and 0.8464 \si{\angstrom} around the origin.
        The off-axis plot is at $\text{z}=-0.0485$ \si{\angstrom} for NEO-MR-SD$_\text{en}$CI but at $\text{z} = 0.0$ \si{\angstrom} for all other methods.
        The NEO-HF maximum protonic density is 23 a.u.}
\label{fig:hnc_gs_density_compare}
\end{figure}

\newpage

{\color{revisions}
\section{Excited vibronic states computed with conventional electronic basis sets}
\begin{table}[h]
\let\TPToverlap=\TPTrlap
\centering
\caption{\color{revisions}Fundamental Vibrational Excitation Energies (in \si{\wn}) for \ce{HCN} and \ce{FHF-} Computed with Conventional Electronic Basis Sets and Revised $\gamma$-def2-QZVP* Electronic Basis Set}
\label{tbl:HCN_and_FHF_with_conventional_basis}
\begin{threeparttable}
{\color{revisions}
\begin{tabularx}{.95\textwidth}{l l Y Y}
    \toprule
    \toprule
    Method & H Basis & Bend & Stretch\\
    \midrule
    \multicolumn{4}{c}{\ce{HCN}$^\tnote{\emph{a}}$}\\
    NEO-MR-SD$_\text{en}$CI$^\tnote{\emph{b}}$  & Dyall v4z &  1512 & 3483\\
    NEO-MR-SD$_\text{en}$CI$^\tnote{\emph{b}}$  & cc-pV5Z   &  1455 & 3498\\
    NEO-MR-SD$_\text{en}$CI$^\tnote{\emph{b}}$  & def2-QZVP &  1528 & 3524\\
    NEO-MR-SD$_\text{en}$CI$^\tnote{\emph{b}}$  & 0.1393-def2-QZVP* & 678 & 2968\\
    Reference$^\tnote{\emph{c}}$                & &    632   &    3152\\
    \multicolumn{4}{c}{\ce{FHF-}$^\tnote{\emph{d}}$}\\
    NEO-MR-SD$_\text{en}$CI$^\tnote{\emph{e}}$  & Dyall v4z &  1878  &  2273\\
    NEO-MR-SD$_\text{en}$CI$^\tnote{\emph{e}}$  & cc-pV5Z   &1842   & 2192 \\
    NEO-MR-SD$_\text{en}$CI$^\tnote{\emph{e}}$  & def2-QZVP &   1881   &    2283\\
    NEO-MR-SD$_\text{en}$CI$^\tnote{\emph{e}}$  & 0.4742-def2-QZVP* & 1475 & 2071\\
    Reference$^\tnote{\emph{f}}$                & &   1298   &    1658\\
    \bottomrule 
\end{tabularx}
\begin{tablenotes}\linespread{1}\footnotesize
    \item[\emph{a}] The C–N distance is 1.153 Å. The H basis function center is 1.065 Å from the C atom.
    \item[\emph{b}] The electronic basis set for C and N is cc-pVTZ. The protonic basis set for H is 8s8p8d8f. The electronic active space is 2 orbitals and 2 electrons.
                    The protonic active space is 16 orbitals and 1 proton.
                   The orbitals are optimized with NEO-CASSCF.           
    \item[\emph{c}] The reference is from an FGH calculation using CCSD and the cc-pVTZ electronic basis set on all atoms.
    \item[\emph{d}] The F-F distance is 2.271 Å. 
    \item[\emph{e}] The electronic basis set for F is aug-cc-pVTZ. The protonic basis set for H is 8s8p8d8f. 
                The electronic active space is 4 orbitals and 4 electrons.
                The protonic active space is 20 orbitals and 1 proton. The orbitals are optimized with NEO-SA-MCSCF averaging over the lowest four vibronic states.
    \item[\emph{f}] The reference is from an FGH calculation using CCSD and the aug-cc-pVTZ electronic basis set on all atoms.
\end{tablenotes}
}
\end{threeparttable}
\end{table}
}

\newpage

{\color{revisions}
\section{Single-reference CI results}
\begin{table}[h]
\let\TPToverlap=\TPTrlap
\centering
\caption{\color{revisions}Fundamental Vibrational Excitation Energies (in \si{\wn}) for \ce{HeHHe+}, \ce{FHF-}, \ce{HCN}, and \ce{HNC} Computed with NEO-SD$_\text{en}$CI\tnote{\emph{a}}}
\label{tbl:neo_cisd_results}
\begin{threeparttable}
{\color{revisions}
\begin{tabularx}{.9\textwidth}{l Y Y}
    \toprule
    \toprule
    Method & Bend & Stretch\\
    \midrule
    \multicolumn{3}{c}{\textbf{\ce{HeHHe+}}}\\
    NEO-SD$_\text{en}$CI\tnote{\emph{b}}    &  699 &    \\
    Reference\tnote{\emph{c}}               &  771 &   \\
    \multicolumn{3}{c}{\textbf{\ce{FHF-}}}\\
    NEO-SD$_\text{en}$CI\tnote{\emph{d}}    & 1260  &  1586 \\
    Reference\tnote{\emph{e}}               & 1298  &  1658\\
    \multicolumn{3}{c}{\textbf{\ce{HCN}}}\\
    NEO-SD$_\text{en}$CI\tnote{\emph{f}}    & 664  &  3143 \\
    Reference\tnote{\emph{g}}               & 632  &  3152 \\
    \multicolumn{3}{c}{\textbf{\ce{HNC}}}\\
    NEO-SD$_\text{en}$CI\tnote{\emph{h}}    &  576 &  3620 \\
    Reference\tnote{\emph{i}}               &  439 &  3547 \\
    \bottomrule 
\end{tabularx}
\begin{tablenotes}\linespread{1}\footnotesize
    \item[\emph{a}] The geometries are the same as those used in                 the tables presented in the main paper.
    \item[\emph{b}] The electronic basis set for He is 6-31G.
                    The electronic and protonic basis sets for H are 0.4404-def2-QZVP* and 8s8pd8d8f, respectively.
    \item[\emph{c}] The reference is from an FGH calculation using FCI and the 6-31G electronic basis set on all atoms.
    \item[\emph{d}] The electronic basis set for F is aug-cc-pVTZ.
                    The electronic and protonic basis sets for H are 0.4742-def2-QZVP* and 8s8p8d8f, respectively.
    \item[\emph{e}] The reference is from an FGH calculation using CCSD and the aug-cc-pVTZ electronic basis set on all atoms.
    \item[\emph{f}] The electronic basis set for C and N is cc-pVTZ.
                    The electronic and protonic basis sets for H are 0.1393-def2-QZVP* and 8s8p8d8f, respectively.
    \item[\emph{g}] The reference is from an FGH calculation using CCSD and the cc-pVTZ electronic basis set on all atoms.
\end{tablenotes}
}
\end{threeparttable}
\end{table}
}

\newpage

{\color{revisions}
\section{Transferability of revised electronic basis sets for molecules with an internal proton}

{\color{revisions}
Table S5 and Figures S9 and S10 show the transferability of the revised electronic basis set for internal hydrogen nuclei. 
}
\begin{table}[htp]
\let\TPToverlap=\TPTrlap
\centering
\caption{\color{revisions}Fundamental Vibrational Excitation Energies (in \si{\wn}) for \ce{FHF-} and \ce{HeHHe+} Computed with the Two Corresponding Revised Electronic Basis Sets\tnote{\emph{a}}}
\label{tbl:fhf_and_hehhe_with_diff_gammas}
\begin{threeparttable}
\color{revisions}
\begin{tabularx}{.9\textwidth}{l Y Y}
    \toprule
    \toprule
    Basis                   & Bend  & Stretch \\
    \midrule
    \multicolumn{3}{c}{\textbf{\ce{FHF-}}\tnote{\emph{a}}}\\
    0.4404-def2-QZVP*       & 1467 & 2089 \\
    0.4742-def2-QZVP*       & 1475 &  2071\\
    \multicolumn{3}{c}{\textbf{\ce{HeHHe+}}\tnote{\emph{b}}}\\
    0.4404-def2-QZVP*       &  786 &      \\
    0.4742-def2-QZVP*       &  809 &      \\
    \bottomrule
\end{tabularx}
\begin{tablenotes}\linespread{1}\footnotesize
    \item[\emph{a}] The F-F distance is 2.271 \si{\angstrom}.
                    The electronic basis set for F is aug-cc-pVTZ.
                    The protonic basis set for H is 8s8p8d8f.
                    The electronic active space is 4 orbitals and 4 electrons.
                    The protonic active space is 20 orbitals and 1 proton.
                    The orbitals are optimized with NEO-SA-MCSCF averaging over the lowest four vibronic states.
    \item[\emph{b}] The He-He distance is 1.8 \si{\angstrom}. 
                    The electronic basis set for He is aug-cc-pVTZ.
                    The protonic basis set for H is 8s8p8d8f.
                    The electronic active space is 6 orbitals and 4 electrons.
                    The protonic active space is 87 orbitals and 1 proton.
                    The orbitals are optimized with NEO-SA-MCSCF averaging over the lowest three vibronic states.
\end{tablenotes}
\end{threeparttable}
\end{table}
}

\begin{figure}[htp]
\centering
\includegraphics[width=.99\linewidth]{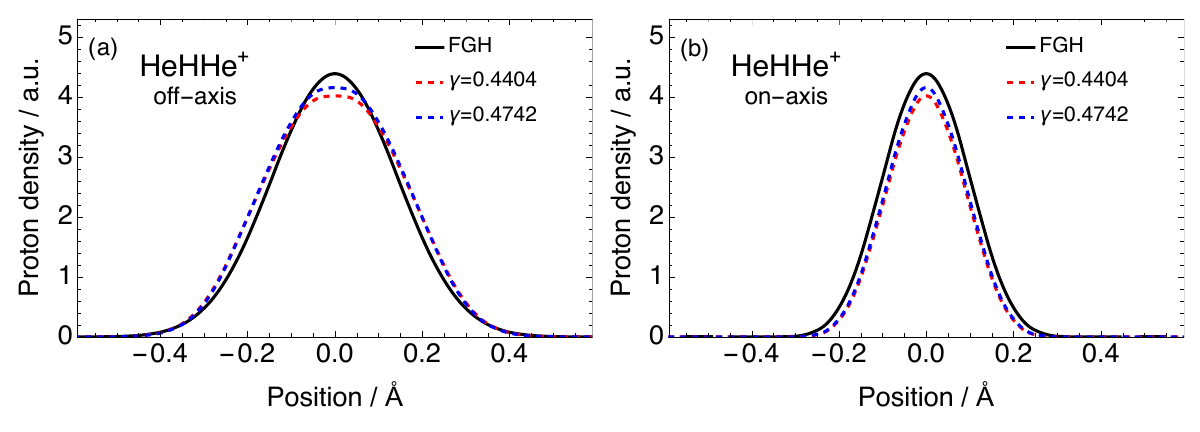}
\caption{\color{revisions}Ground state protonic density (a) off-axis and (b) on-axis for \ce{HeHHe+} computed with the NEO-MR-SDT$_\text{een}$CI method compared to reference FGH results.
The distance between the fixed He atoms is 1.8 \si{\angstrom} with the midpoint at the origin.
The electronic basis set for He is 6-31G.
The electronic basis set for H is $\gamma$-def2-QZVP*, where the value of $\gamma$ is indicated in the legend.
The protonic basis set for H is 8s8p8d8f and is placed at the origin.
The electronic active space is 6 orbitals and 4 electrons, and the protonic active space is 87 orbitals and 1 proton.
The reference FGH calculation uses FCI and the 6-31G electronic basis set for all atoms.
The grid spans the range of -0.8203 to 0.8750 \si{\angstrom} around the origin.}
\label{fig:hehhe_fhfbasis_gs_density_compare}
\end{figure}

\begin{figure}[htp]
\centering
\includegraphics[width=.99\linewidth]{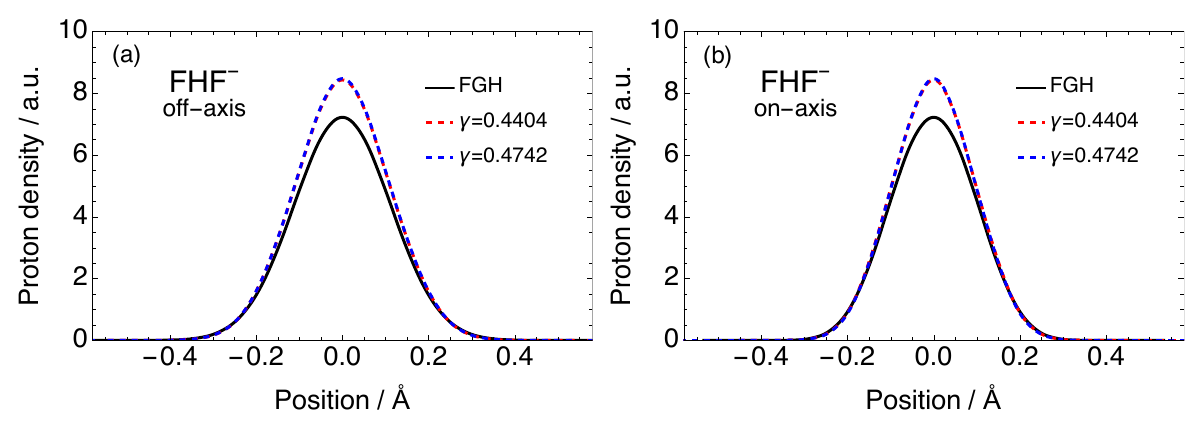}
\caption{\color{revisions}Ground state protonic density (a) off-axis and (b) on-axis for \ce{FHF-} computed with the NEO-MR-SD$_\text{en}$CI method compared to the reference FGH results. 
The distance between the fixed F atoms is 2.271  \si{\angstrom} with the midpoint at the origin. 
The electronic basis set for F is aug-cc-pVTZ.
The electronic basis set for H is $\gamma$-def2-QZVP*, where the value of $\gamma$ is indicated in the legend.
The protonic basis set for H is 8s8p8d8f and is placed at the origin.
The electronic active space is 4 orbitals and 4 electrons, and 
the protonic active space is 20 orbitals and 1 proton.
The reference FGH calculation uses CCSD and the aug-cc-pVTZ electronic basis set for all atoms.
The grid spans the range of -0.7275 to 0.7760 \si{\angstrom} around the origin.}
\label{fig:fhf_hehhebasis_gs_density_compare}
\end{figure}

\newpage

\bibliography{supporting_information}